\documentclass[aps,amsmath,twocolumn,amssymb,floatfix,showpacs,superscriptaddress,nofootinbib,longbibliography]{revtex4-1}
\usepackage{MnSymbol}
\usepackage{braket}
\usepackage[dvipsnames]{xcolor}
\usepackage{float}
\usepackage{subfigure}
\usepackage{tikz}
\usepackage[colorlinks=true,linktoc=page,linkcolor=Purple,citecolor=RedViolet,urlcolor=Magenta]{hyperref}
\usepackage{amssymb} 
\usepackage{amsfonts}

\mathchardef\mhyphen="2D 

\newcommand{\ie}{{ i.e.,\,\,}}
\newcommand{\eg}{{e.g.,~}}

\newcommand\bea{\begin{eqnarray}}
	\newcommand\eea{\end{eqnarray}}
\newcommand\beq{\begin{equation}}  
	\newcommand\eeq{\end{equation}}

\usepackage{float}

\usepackage[normalem]{ulem}
\definecolor{lime}{HTML}{A6CE39}
\usepackage{sidecap,tikz}
\usepackage{amsmath, amssymb}
\DeclareRobustCommand{\orcidicon}{\hspace{-1.0mm}
	\begin{tikzpicture}
		\draw[lime, fill=lime] (0.0,0.0) 
		circle [radius=0.15] 
		node[white] {{\fontfamily{qag}\selectfont \tiny \,ID}};
		\draw[white, fill=white] (-0.0525,0.095) 
		circle [radius=0.007];
	\end{tikzpicture}
	\hspace{-3.0mm}
}
\foreach \x in {A, ..., Z}{\expandafter\xdef\csname orcid\x\endcsname{\noexpand\href{https://orcid.org/\csname orcidauthor\x\endcsname}{\noexpand\orcidicon}}
}

\AtBeginDocument{%
	\newwrite\bibnotes
	\def\bibnotesext{Notes.bib}
	\immediate\openout\bibnotes=\jobname\bibnotesext
	\immediate\write\bibnotes{@CONTROL{REVTEX41Control}}
	\immediate\write\bibnotes{@CONTROL{%
			apsrev41Control,author="08",editor="1",pages="1",title="1",year="1"}}
	\if@filesw
	\immediate\write\@auxout{\string\citation{apsrev41Control}}%
	\fi
}%

\begin{document}
	
	\title{Engineering second order topological superconductor hosting tunable Majorana corner modes in magnet/$d$-wave superconductor hybrid platform}  
	
	\author{Minakshi Subhadarshini\orcidA{}}
	\email{minakshi.s@iopb.res.in}
	\affiliation{Institute of Physics, Sachivalaya Marg, Bhubaneswar-751005, India}
	\affiliation{Homi Bhabha National Institute, Training School Complex, Anushakti Nagar, Mumbai 400094, India}
	\author{Archana Mishra\orcidB{}}
	\email{amishra@physics.du.ac.in}
	\affiliation{Department of Physics and Astrophysics, University of Delhi, Delhi-110007}
	\author{Arijit Saha\orcidD{}}
	\email{arijit@iopb.res.in}
	\affiliation{Institute of Physics, Sachivalaya Marg, Bhubaneswar-751005, India}
	\affiliation{Homi Bhabha National Institute, Training School Complex, Anushakti Nagar, Mumbai 400094, India}

\begin{abstract}
We theoretically study the  noncollinear magnetic texture effect 
on second-order topological superconductor (SOTSC) phase generated in unconventional $d$-wave superconductors and two-dimensional (2D)
quantum spin Hall insulators (QSHI).
While the interplay of the $d$-wave superconductor and QSHI has been studied as a platform to realize  Majorana corner modes (MCMs), we show that the addition of the spin texture enables the tunability of these MCMs. Each corner of this hybrid system can host one or two Majorana modes depending on the system parameters, in particular, exchange strength and pitch vector of the spin texture. To characterize the higher order bulk topology, we compute the quadrupolar winding number, which directly corresponds to the number of MCMs acquiring a value of one for four corner modes and two for eight corner modes. We investigate and show the close resemblance in the topological phase diagrams obtained from the low energy effective Hamiltonian that reveals an emergent in-plane Zeeman field and spin-orbit coupling induced by the spin texture, and the real space tight binding lattice model. The microscopic pairing mechanism responsible for the appearance of SOTSC phase is investigated via an effective bulk pairing analysis, while a low-energy edge theory captures the mechanism behind tunability of MCMs. Our result paves the way for realizing SOTC with multiple MCMs which can be tuned via system parameters. 
\end{abstract}

\maketitle

\section{Introduction}
Majorana Zero modes (MZMs) in topological superconductor (TSCs) have emerged as a central focus of condensed matter research due to their exotic non-Abelian braiding statistics and their potential applications as building blocks for fault-tolerant topological quantum computation \cite{Kitaev2001,Cheng_2011,Ivanov2001,SDSarma2008,beenakker2013search,Alicea_2012,Leijnse_2012,Kitaev2009}. In conventional first-order TSCs, MZMs manifest as gapless boundary states at $(d-1)$-dimensional edges of a $d$-dimensional system. The advent of higher-order topological systems has significantly broadened this understanding of bulk-boundary correspondence, introducing higher-order topological superconductors (HOTSCs), which host gapless Majorana states at $(d-n)$-dimensional boundaries, where $n\geq 2$~\cite{Geier_2018,Zhong_2017,Khalaf_2018,zhong2018,Volovik2010}. For instance, a second-order topological superconductor (SOTSC) in two dimensions (2D) can support zero-dimensional (0D) localized Majorana corner modes (MCMs) at the corners of 2D domain.
	
Over the past decade, a substantial progress has been made in engineering first order TSC phases using heterostructures composed of materials with strong spin-orbit coupling (SOC), such as topological insulators, semiconductor thin films, and nanowires with proximity-induced superconductivity~\cite{Fu2008, Lutchyn2010, Oreg2010, Mourik2012, Das2012}. Experimental evidence of MZMs in such platforms has reinforced the feasibility of topological quantum computation~\cite{Mourik2012, Das2012, Albrecht2016}. In these systems, MZMs typically appear at one-dimensional (1D) edge or 2D vortex cores, where the topological superconducting gap in the bulk spectrum undergoes a sign change~\cite{Fu2008}.
	
A particularly promising approach for realizing TSC phases involves magnet/superconductor hybrid (MSH) structures, where the interaction between magnetic impurity spins and conventional $s$-wave superconductivity gives rise to Yu-Shiba-Rusinov (YSR) states within the superconducting gap~\cite{Felix2013,AliYazdani2013,DanielLoss2013,PascalSimon2013,MFranz2013,Eugene2013,Felix2014,TeemuOjanen2014,MFranz2014,Rajiv2015,Sarma2015,
Hoffman2016,Jens2016,Tewari2016,PascalSimon2017,Simon2017,Theiler2019,Cristian2019,Mashkoori2019,Menard2019,Pradhan2020,Teixeira2020,
Alexander2020,Perrin2021,Nicholas2020,Chatterjee_2024a,Chatterjee_2024b,Mondal2023,Jelena2016,Balatsky22016,subhadarshini2025,RACHEL20251,
LoConte2024}. The hybridization of YSR states can form an emergent Shiba band within the parent superconducting gap that effectively mimics a $p$-wave superconductor within the induced minigap, enabling the emergence of MZMs. Experimental realizations of MZMs and YSR states in magnetic impurity chains deposited on superconducting substrates~\cite{Eigler1997,Yazdani1999,Yazdani2015,Wiesendanger2021,Beck2021,Wang2021,Schneider2022,Richard2022,Wiesendanger2022,Yacoby2023,Soldini2023} underscore the potential of this approach. Furthermore, noncollinear magnetic textures in proximity with conventional $s$-wave and unconventional $d$-wave superconductors have been shown to give rise to one-dimensional (1D) Majorana flat edge modes (MFEMs) at the boundary of 2D Shiba lattice~\cite{Chatterjee_2024a,Subhadarshini2024,Neupert2016}. 
	
In the context of HOTSCs, theoretical proposals predict the possibility of engineering SOTSC in 2D heterostructures, offering a promising route to realize MCMs. For instance, a monolayer of 
$\text{FeTe}_{1-x}\text{Se}_{x}$, a superconducting quantum spin Hall material coupled with a monolayer of FeTe, which exhibits bicollinear antiferromagnetism, has been predicted to host MCMs~\cite{DasSharma_2019}. Similarly, a 2D electron gas with Rashba spin-orbit coupling (SOC), proximity-induced $s$-wave superconductivity, and an in-plane Zeeman field has also been shown to support MCMs~\cite{Dloss_2019}. Recent studies have further suggested that intrinsic noncollinear spin textures, in place of an external Zeeman field and QSHI, can give rise to SOTSC, generating significant interest in this mechanism~\cite{Chatterjee_2024b}.	Moreover, topological insulators (TIs) and their higher-order counterparts have been shown to generate MCMs in 2D and Majorana hinge modes in three-dimensional (3D) superconductor (SC) heterostructures~\cite{Ghosh_2021}. High-temperature platforms based on $d$-wave SCs have also been explored for realizing MCMs, including those coupled to QSHIs and Zeeman fields~\cite{zhong2018}, as well as in altermagnet heterostructures~\cite{Cheng-Cheng_2023}. Furthermore, SOTSC has been predicted in magnetic topological insulators proximitized with high-temperature superconductors with both $d$-wave and $s_{\pm}$-wave pairing symmetries~\cite{Franco_2018}. Majorana Kramers pairs have also been proposed to emerge in 2D TIs coupled to Fe-based superconductors~\cite{Cheng-Cheng_2018}. However, the potential for generating MCMs in $d$-wave superconductors coupled with noncollinear magnetic textures and QSHIs remains unexplored, presenting a compelling direction for further research.

    	\begin{figure}
	\centering
	\subfigure{\includegraphics[width=0.5\textwidth]{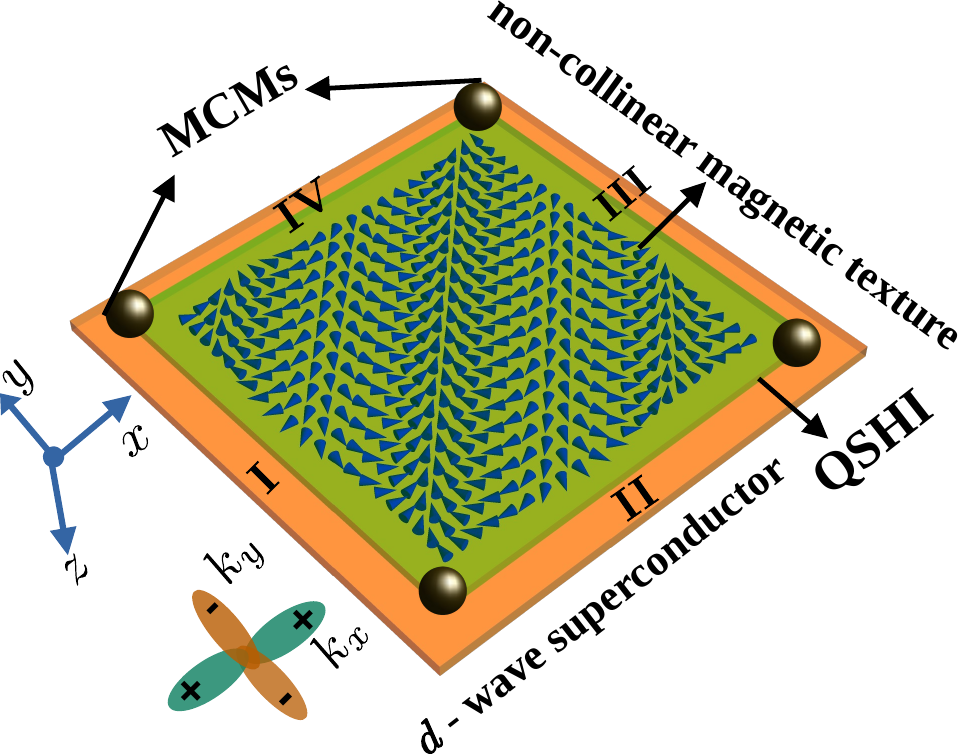}}
	\caption{Schematic diagram of our 2D heterostructure consisting of an unconventional $d$-wave SC shown in orange at the bottom, with $d$-wave pairing lies in the $x$-$y$ plane. A noncollinear spin texture, represented by blue arrows, is deposited on the surface of the $d$-wave SC. On top of this spin texture, a layer of quantum spin Hall insulator (QSHI), depicted in green, is deposited resulting in a sandwich-like configuration.  The four edges of the 2D heterostructure are labeled as I, II, III, and IV. The system hosts localized MCMs, indicated by four black spheres at the corners of the heterostructure.}
	\label{Fig1}
\end{figure}
	
	Motivated by these facts, in this article, we pose the following key questions in this work: (a) Can MCMs be realized in a 2D heterostructure combining a $d$-wave SC, a QSHI, and a noncollinear magnetic texture? (b) Are the resulting MCMs tunable in number by varying the external or internal parameters? (c) How to construct the corresponding edge theory and MCMs solution for an insight into the tunability of the MCMs?
	
	To address these questions, we propose a scheme for engineering SOTSC hosting MCMs in a hybrid platform consisting of a QSHI, a noncollinear magnetic texture, and a $d$-wave superconductor. Unlike previous approaches relying on Rashba SOC and Zeeman fields, our lattice model employs the intrinsic pairing structure of $d$-wave superconductivity and its interplay with magnetic textures to stabilize and tune the MCMs that appear in SOTSC phase. We further derive an effective momentum-space Hamiltonian via a unitary transformation and explore the nature of the pairing symmetries responsible for the stabilization of SOTSC phase. Our effective model provides analytical insights by calculating the low-energy edge theory and MCMs solution that govern the tunability of MCMs. We further characterize the corresponding SOTSC phase by appropriate topological invariant, the quadrupolar winding number, that determines the number of MCMs at the corners of our 2D domain.
	
	The remainder of this paper is organized as follows. In Sec.~\ref{Sec:II}, we introduce the real-space lattice Hamiltonian (some limits of it are highlighted in Appendix~\ref{Appendix-A}) corresponding to our setup and demonstrate the emergence of four and eight MCMs in our 2D heterostructure. In Sec.~\ref{Sec:III}, we derive the effective momentum space Hamiltonian, determine the edge spectrum, and analyze the pairing symmetries responsible for the emergence of MCMs. The detailed calculations are provided in Appendices \ref{Appendix-B} and \ref{Appendix-C}. Sec.~\ref{Sec:IV} is devoted to the topological characterization of the SOTSC phase employing both the full lattice model and the effective low-energy Hamiltonian. In Sec.~\ref{Sec:V}, we construct the low-energy edge theory to derive analytical solutions for MCMs, with detailed derivations in Appendices \ref{Appendix-D} and \ref{Appendix-E}. We discuss the effect of Rashba SOC in Appendix~\ref{Appendix-F}. Finally, we summarize and conclude our paper in Sec.~\ref{Sec:VI}. In Appendix~\ref{Appendix-G}, we discuss the effect of conical spin texture and in Appendix~\ref{Appendix-H}, we show the emergence of MCMs in traingular geometry.

\section{Lattice Model Hamiltonian }{\label{Sec:II}}
In this section, we introduce the lattice Hamiltonian governing our system, which comprises of a $d$-wave SC, a QSHI, and a noncollinear spin texture. The interplay among these layers plays a crucial role in stabilizing different topological phases. By systematically tuning the parameters, we explore the conditions under which  first-order as well as second-order zero-energy modes emerge. To further investigate the tunability of the second-order modes, we numerically compute the eigenvalue spectrum, construct phase diagrams representing the number of MCMs, and analyze the LDOS at $E = 0$ to visualize the spatial localization of zero-energy modes.

The full real space lattice Hamiltonian of the system is given by~\cite{Chatterjee_2024a,Chatterjee_2024b,Subhadarshini2024}:
\begin{equation}
	\begin{aligned}
		H = \sum_{i,j} c_{i,j}^{\dagger} \bigg[ & \left\{\epsilon_0 \Gamma_1+J(\Gamma_3\cos \phi_{i,j}+\Gamma_4 \sin \phi_{i,j})\right\}c_{i,j} \\
		& -\left(t \Gamma_1+i\lambda_x\Gamma_5-\Delta_0\Gamma_2 \right)c_{i+1,j} \\
		& -\left(t\Gamma_1+i\lambda_y\Gamma_6+\Delta_0\Gamma_2\right)c_{i,j+1} \bigg] + h.c.\ ,
		\label{eq:HamiltonianExact}
	\end{aligned}
\end{equation}
where, $(i, j)$ denote lattice sites along the $x$- and $y$-directions, respectively as shown in Fig.~\ref{Fig1}. The $\Gamma$ matrices, which encode the system’s internal degrees of freedom, are defined as:
\begin{align*}  
	\Gamma_1 &= \tau_z\sigma_zs_0, & \Gamma_2 &= \tau_x\sigma_0s_0, & \Gamma_3 &= \tau_0\sigma_0s_x, \\  
	\Gamma_4 &= \tau_0\sigma_0s_y, & \Gamma_5 &= \tau_z\sigma_xs_z, & \Gamma_6 &= \tau_z\sigma_y s_0.  
\end{align*}  
Here, $\tau$, $\sigma$, and $s$ are Pauli matrices acting on the particle-hole, orbital, and spin degrees of freedom, respectively and the Hamiltonian is written in the Nambus basis $c_{i\in  (x,y)}=\left(c_{ia\uparrow},c_{ia\downarrow},c_{ib\uparrow},c_{ib\downarrow},-c_{ia\downarrow}^{\dagger},c_{ia\uparrow}^{\dagger},-c_{ib\downarrow}^{\dagger},c_{ib\uparrow}^{\dagger}\right)^T$
where $T$ denotes the transpose operation.
The parameters $\epsilon_0$, $\Delta_0$, $t$, $\lambda_{x,y}$, and $J$ represent the staggered mass term in the QSHI, $d$-wave superconducting order parameter, nearest-neighbor hopping amplitude, SOC strength in $x$ and $y$- direction, and local exchange interaction between the magnetic adatom spins and superconducting electrons, respectively. Here, $\phi_{i,j}$ denotes the angle between two magnetic impurities located at two neighbouring sites. In our analysis, all other energy scales are scaled in terms of the hopping parameter $t$ and $t=1$. 

The magnetic impurities are treated as classical spins and are represented by a unit spin vector~\cite{Shiba1968,Felix2013}:
\begin{equation}
	\textbf{S(r)}=S(\sin \theta_{r} \cos \phi_{r}, \sin \theta_{r} \sin \phi_{r}, \cos \theta_{r})\ ,
\end{equation}
where, $\theta_{r}$ and $\phi_{r}$ denote the polar and azimuthal angles at position $r$ and $S$ represents the magnitude of the spins in the spin texture which, from hereon, we consider to be unity. 
The spin texture is assumed to vary along both $x$ and $y$ directions, parameterized as $\theta_r=\pi/2$ and $\phi_{r} = g_xx + g_yy$~\cite{Chatterjee_2024a,Chatterjee_2024b,Subhadarshini2024}. Here, $g_x$ and $g_y$ define the pitch vectors governing the periodicity of the spin modulation along the $x$- and $y$-directions, respectively. 

To gain deeper insight into the emergence of various topological phases, we analyze four key limiting cases of our Hamiltonian:

\textbf{Case-I:} When $J = \Delta_0 = 0$, the Hamiltonian in Eq.~(\ref{eq:HamiltonianExact}) reduces to the Bernevig-Hughes-Zhang (BHZ) model for a 2D QSHI, which hosts first-order topological 
phase with helical edge modes and preserves time-reversal symmetry (TRS)~\cite{science-Bernevig,science-zhang}.

\textbf{Case-II:} Considering only the interactions between the $d$-wave SC and the spin texture ($\lambda_x = \lambda_y = 0$), multiple Majorana flat edge modes (MFEMs) emerge in this system \cite{Subhadarshini2024}.

\textbf{Case-III:} The interactions between the QSHI and the non-collinear spin texture in the absence of the suprconductor ($\Delta_0 = 0$ and $J, g_x, g_y \neq 0$) breaks the TRS which results in gapped helical edge states (see Appendix \ref{Appendix-A}) for details.

\textbf{Case-IV:} Coupling of the QSHI to the $d$-wave SC without any spin texture ($J=0$) induces higher order topological modes with the emergence of eight Majorana corner modes (MCMs) at the 
corners of the 2D square lattice~\cite{zhong2018} (also, see Appendix~\ref{Appendix-A} for details).

The above limiting cases build the foundation for understanding the complete phase diagram in the composite (three layer) system of the QSHI, $d$-wave SC and spin texture as schematically 
depicted in Fig.~\ref{Fig1} and formally described by Eq.~(\ref{eq:HamiltonianExact}). This, along with establishing a feasible path for realizing the SOTSC phase, allows tuning of the MCMs using 
the spin texture. 

\begin{figure}
		\centering
		\subfigure{\includegraphics[width=0.5\textwidth]{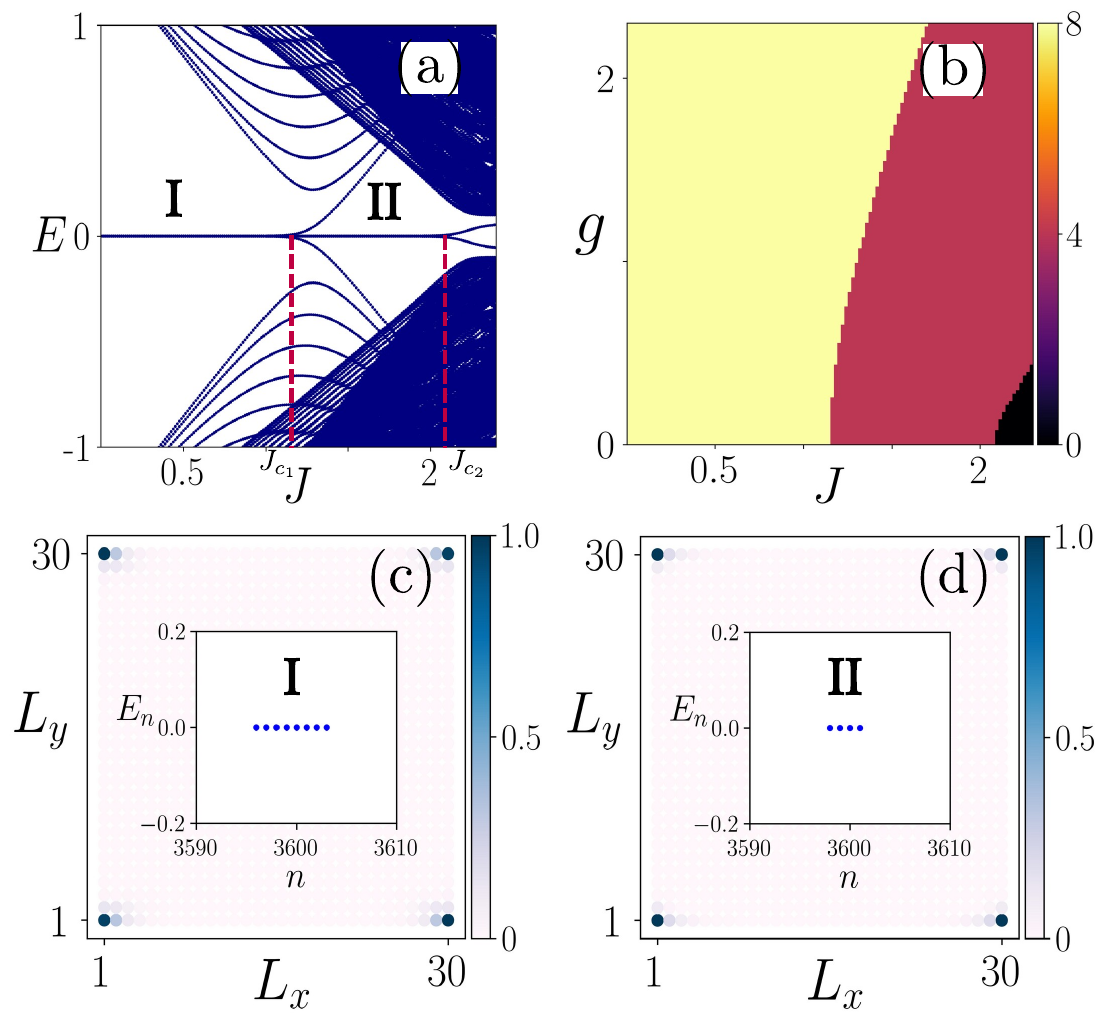}}
		\caption{Panel (a) depicts the energy eigenvalue spectrum, obtained via OBC, as a function of the exchange interaction $J$, with fixed model parameters $g_x = g_y = g = 0.1$, $\Delta_0 = 0.5t$, and $\lambda_x = \lambda_y = \lambda = \epsilon_0=t$. The red dotted lines indicates the two critical values of $J$ represented as $J_{c_1}$ and $J_{c_2}$. For the former case topological phase transition takes place from one SOTSC phase (8 MCMs) to second SOTSC phase (4 MCMs) and the latter showcase the transition from topological to trivial phase. Region I corresponds to the presence of 8 MCMs, as illustrated in the inset of panel (c) choosing $J = 0.8t$, while region II denotes the presence of 4 MCMs, as shown in the inset of  panel (d) for $J = 1.5t$. In panels (c) and (d) LDOS spectra of the MCMs are depicted in the $L_x-L_y$ plane. The insets of (c) and (d) represent the eigenvalue spectrum which we have zoomed around zero energy. In panel (b), we present the phase diagram for number of MCMs in the $g$-$J$ plane, where the yellow region corresponds to 8 MCMs and the pink region to 4 MCMs. All other model parameters remains same as mentioned above. Here we choose a finite size system with $30\times 30$ lattice sites.} 
		\label{Fig2}
	\end{figure}

To further investigate the second-order topological phase, we compute the eigenvalue spectrum employing open boundary condition (OBC), in both $x$ and $y$ directions, as a function of $J$ as shown in Fig.~\ref{Fig2}(a). Two distinct band-closing points signal transitions between different topological phases: the first transition reflects a change in the number of MCMs, while the second marks the crossover from a topological phase with four MCMs to a trivial gapped SC phase. These transitions occur at two critical values, $J_{c_1}$ and $J_{c_2}$ as indicated by the dotted lines in Fig.~\ref{Fig2}(a). The topological phase associated with the 8-MCMs appear in region I, while region II denotes the SOTSC phase with 4-MCMs, as illustrated in the inset of Figs.~\ref{Fig2}(c) and \ref{Fig2}(d), respectively. In both the cases, the Majorana zero modes are localized at the corners of the 2D domain, consistent with the LDOS calculations (see Figs.~\ref{Fig2}(c) and \ref{Fig2}(d)).

The phase diagram in Fig.~\ref{Fig2}(b) maps the number of MCMs within the $g$-$J$ parameter space, revealing three distinct regions: the yellow region represents the SOTSC phase with 8-MCMs while the pink region corresponds to the same with 4-MCMs. Then the black region indicates a trivial gapped phase with no protected modes. Note that, when $J=0$, the SOTSC phase only contains 8 MCMs which is consistent with Ref.~\cite{zhong2018}.

\section{Effective Low-Energy Continuum Model}  \label{Sec:III}
To develop a deeper analytical understanding of our findings, we derive a low-energy continuum version of the Hamiltonian described by Eq.~(\ref{eq:HamiltonianExact}). This effective model allows 
one to identify the underlying pairing mechanism responsible for the emergence of MCMs as well as to construct the low energy edge theory described later. Our analysis is conducted within a mixed-space framework, incorporating both the real and momentum space, which starts from the following low-energy Hamiltonian~\cite{Chatterjee_2024b,Subhadarshini2024}:  
\begin{equation}
	\begin{aligned}
		H_{L} =  \xi_k \Gamma_1 + \Delta_d\Gamma_2 
		+ 2\lambda_x k_x a\Gamma_5 + 2\lambda_y k_y a\Gamma_6 + J\mathbf{S(r)} \cdot \mathbf{s}\ ,
	\end{aligned}
	\label{eq:Low energy H}
\end{equation}  

where, $\xi_k = \epsilon_0 - 4t + ta^{2}(k_x^2 + k_y^2)$ represents the kinetic energy and $\Delta_d = \Delta_0 a^{2}(k_x^2 - k_y^2)$ denotes the intrinsic $d$-wave SC pairing~\cite{zhong2018}. The terms $2\lambda_x k_x a \Gamma_5 + 2\lambda_y a k_y\Gamma_6$ correspond to the intrinsic SOC inherent to the QSHI~\cite{science-Bernevig}. The last term, $J\mathbf{S(r)} \cdot \mathbf{s}$, describes the exchange coupling due to the spin texture in real space. Here, $a$ corresponds to the lattice spacing and we have considered $a=1$ throughout our article. Also, any other length scales
are scaled with respect to $a$. Note that, the first four terms are written in momentum space, while the exchange coupling term reflects the influence of the real-space spin texture, coupled with SC electrons.
\subsection{Effective low-energy Hamiltonian}
The real-space dependence in the Hamiltonian $H_L$ can be eliminated by a unitary trasnformation $U = \tau_0\sigma_0 e^{-i\frac{\phi}{2} s_z}$~\cite{DLoss_2022,Chatterjee_2024b,Subhadarshini2024} and the resultant effective low energy Hamiltonian can be written as (see Appendix~\ref{Appendix-B} for detailed derivation) 
 
\begin{equation}
	\begin{aligned}
		H_{\text{eff}} = & \xi_{k_\text{eff}} \Gamma_1 + \Delta_{d_\text{eff}} + J\Gamma_3  
		+ 2\lambda_x k_x \Gamma_5 + 2\lambda_y k_y \Gamma_6 \\
		& - \frac{t}{2} \mathbf{g} \cdot \mathbf{k} \Gamma_7 
		- \lambda_x g_x \Gamma_8 - \lambda_y g_y \Gamma_9 \ ,  
	\end{aligned}
	\label{eq:EffH1}
\end{equation}  
where, $\xi_{k_{\text{eff}}} = \left[ \xi_k - \frac{t}{4}(g_x^2 + g_y^2) \right] \Gamma_1$ and $\Delta_{d_\text{eff}} = \Delta_d \Gamma_2 + \frac{\Delta_0}{2} (g_y k_y - g_x k_x) \Gamma_{10}$.
The additional $\Gamma$ matrices are defined as $\Gamma_7 = \tau_z\sigma_z s_z,  \Gamma_8 = \tau_z\sigma_x s_0,  \Gamma_9 = \tau_z\sigma_y s_z,  \Gamma_{10} = \tau_x\sigma_0 s_z.$ 

From Eq.~(\ref{eq:EffH1}), it is evident that the kinetic energy undergoes renormalization by a constant factor due to the spin texture. The intrinsic $d$-wave pairing term, $\Delta_d$, acquires an emergent contribution of the form $(p_x - p_y)$, represented by the term $\frac{\Delta_0}{2} (g_y k_y - g_x k_x)$ in $\Delta_{d_\text{eff}}$. In addition to the inherent SOC terms, ($\lambda_y k_y$ and $\lambda_x k_x$), there are extra SOC-like terms of the form $g_x k_x + g_y k_y$ that appear due to the underlying spin texture configuration. Furthermore, an effective Zeeman field proportional to $J$ emerges from the same mechanism. Despite these analytical modifications, the underlying physics remains consistent with the original full lattice model. Throughout our analysis, we assume a symmetric pitch vector, such that $g_x = g_y = g$. 

Here, we discuss the symmetry properties of different limits of $H_{\text{eff}}$. When there is no spin texture ($J = 0, g_x = g_y = 0$), the Hamiltonian in Eq.~(\ref{eq:EffH1}) preserves time-reversal symmetry (TRS), chiral symmetry ($S$), and particle--hole symmetry ($C$). In this case, TRS is given by $T = i \tau_0 \sigma_0 s_y \kappa$ with $\kappa$ is the complex conjugation operator and 
$T H_{\text{eff}}(k) T^{-1} = H_{\text{eff}}(-k)$, particle-hole symmetry by $C = \tau_y \sigma_0 s_y \kappa$ with $C H_{\text{eff}}(k) C^{-1} = - H_{\text{eff}}(-k)$, and chiral symmetry by $S = \tau_y \sigma_0 s_z$ with $S H_{\text{eff}}(k) S^{-1} = - H_{\text{eff}}(k)$. When $J$, $g_x$, and $g_y$ are turned on, TRS is broken by the in-plane Zeeman term, but the system still respects a pseudo TRS $T' = \tau_0 \sigma_0 s_x \kappa$ satisfying $T'^{-1} H_{\text{eff}}(k) T' = H_{\text{eff}}(-k)$, thereby changing the symmetry class from DIII to BDI.

To regularize the effective model on the lattice, we use the approximations:  
\[
k_y \approx \sin k_y, \quad k_y^2 \approx 2(1 - \cos k_y).
\] Substituting this in Eq.~(\ref{eq:EffH1}), the lattice regularized Hamiltonian becomes:
\begin{equation}
	\begin{aligned}
		H_{\text{lat}} = & \left[\epsilon_0 - 2t \cos(k_x) - 2t \cos(k_y)\right] \Gamma_1 - \lambda_x g_x \Gamma_8- \lambda_y g_y \Gamma_9\\
		&-\frac{t}{2} (g_x \sin k_x+g_y \sin k_y)\Gamma_7 
		 + J\Gamma_3  
		+ 2\lambda_x \sin k_x \Gamma_5\\
		& + 2\lambda_y \sin k_y \Gamma_6
		+ \Delta_0 [(2\cos k_y - 2\cos k_x)\Gamma_2\\
		& + \frac{1}{2}(g_y \sin k_y - g_x \sin k_x)\Gamma_{10}]  \ ,
	\end{aligned}
	\label{eq:H_latt}	
\end{equation}
Using the lattice regularized model described in Eq.~(\ref{eq:H_latt}), we analyze the edge states of our hybrid system. In order to do this, we consider a quasi-1D nanowire configuration by imposing OBC along the $y$-direction and periodic boundary condition (PBC) along the $x$-direction.

\begin{figure}
	\centering
	\includegraphics[width=0.48\textwidth]{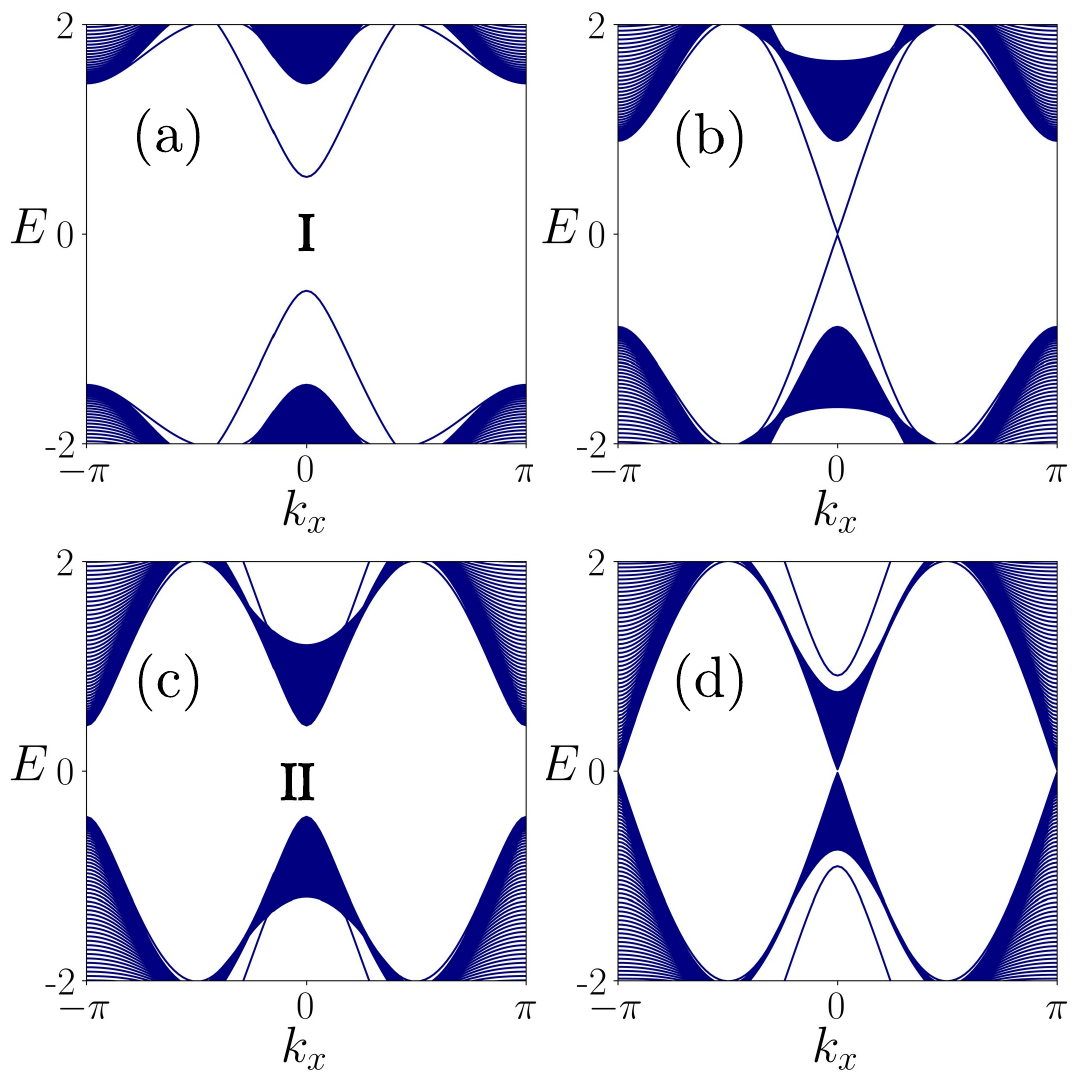}
	\caption{We depict the edge state spectrum obtained under OBC along $y$ and PBC along $x$ direction. Panel (a) and (c) showcase the gapped edge states in SOTSC phases with 8 and 4 MCMs for $J = 0.8t$ and $J = 1.8t$, respectively. Panels (b) and (d) display the gapless edge states at the topological transition points obtained for $J_{c_1} = 1.3t$ and $J_{c_2} = 2.1t$. The other model parameters are chosen as $\Delta_0 = 0.5t$, $g = 0.05$, $t = 1$, $\lambda = t$, and $\epsilon_0 = t$. Here, we consider finite size of $100$ latice sites along $y$-direction.}
	\label{Fig3}
\end{figure}

In Fig.~\ref{Fig3}(a) and Fig.~\ref{Fig3}(c), we show the energy spectra of the edges in two distinct topological phases, corresponding to regions I and II as introduced in Sec.~\ref{Sec:II}. The appearance of SOTSC phase hosting 8 and 4 MCMs for $J = 0.8t$ and $J = 1.8t$, respectively, confirms that the edge spectrum remains gapped in these phases, consistent with the requirement for the emergence of corner modes. In addition, Fig.~\ref{Fig3}(b) and Fig.~\ref{Fig3}(d) depict the topological phase transitions in the parameter space. The first transition, where the number of MCMs reduces from 8 to 4, occurs at the critical point $J_{c_1} = 1.3t$. The second one, which marks the transition from a topological to a trivial phase, takes place at $J_{c_2} = 2.1t$. At both the critical points, the edge states become gapless, signaling the closing and reopening of the bulk gap-a well-known feature of topological phase transitions.

\subsection{Effective Pairing}  
We now examine the effective pairing structure in the bulk based on the low-energy effective Hamiltonian $H_{\text{eff}}$ (see Appendix~\ref{Appendix-C} for detailed derivation). Under a duality transformation using $U_d=\bar{\tau} \sigma_0 s_0$, the dual Hamiltonian can be written as:  
\begin{equation}
	H_D = \begin{pmatrix} 
		\epsilon_D + J \sigma_0 s_x & \Delta_D \\ 
		\Delta_D & -\epsilon_D + J \sigma_0 s_x 
	\end{pmatrix}	\ ,
\end{equation}  
where, $\epsilon_D = \frac{\Delta_0}{2}g(k_y-k_x) \sigma_0 s_z$ represents the dual kinetic energy and $\Delta_D$, the effective pairing. Note that, $\Delta_D$ decomposes into two forms as $\Delta_D = \Delta_s + \Delta_p$. Here,
\[
\Delta_s = (\epsilon_0 - 4t - \frac{t}{2} g^2) \sigma_z s_0 - \lambda g (\sigma_x s_0 + \sigma_y s_z)\ ,
\]  
Here $\Delta_s$ represents the effective $s$-wave pairing and $\Delta_p$, the effective $p$-wave pairing. Note that, $\Delta_p$ consists of a $p_x + i p_y$ type pairing emerging from QSHI and a 
$p_x + p_y$ type pairing that arises from the spin texture. Hence,
\[
\Delta_p = 2\lambda (k_x \sigma_x s_z + k_y \sigma_y s_0) - \frac{t}{2} (\textbf{g.k}) \sigma_z s_z\ .
\]  
This effective pairing $\Delta_D$ stabilizes the SOTSC phase anchoring MCMs in the 2D system.  

\section{Topological Characterization} \label{Sec:IV}  
From the bulk boundary correspondance, the different topological phases characterized by the number of MCMs can be classified by introducing bulk topological invariants that quantify these phases. The appropriate topoloigcal invariant for this system is the quadrupolar winding number which is represented as \( N_{xy} \).  The bulk quadrupole moment alone is insufficient to capture the configurations with multiple MCMs per corner~\cite{Benalcazar_2022}. The quadrupolar winding number provides a more refined characterization, taking discrete values: \( N_{xy} = 1 \) for a single MCM per corner and \( N_{xy} = 2 \) when two MCMs appear at each corner.  

Our system preserves chiral symmetry, which imposes the constraint  $S^{\dagger} H S = -H$,
where \( S \) is the chiral symmetry operator. In a basis where \( S = \tau_y \sigma_0 s_z \), the Hamiltonian can be brought into a block off-diagonal form, $H = U_c H U_c^\dagger =  
\begin{bmatrix}  
0 & h \\  
h^{\dagger} & 0  
\end{bmatrix}, $ 
where \( U_c \) is a unitary matrix composed of the chiral basis eigenvectors. This structure partitions the system into two sublattice degrees of freedom, labeled by \( A \) and \( B \), corresponding to the eigenvalues \( +1 \) and \( -1 \) of \( S \), respectively. Therefore, the eigenvectors of \( H \) can be written as $\ket{\psi_n} = [\ket{\psi_n^A}~\ket{\psi_n^B} ]^T, $ 
where \( \ket{\psi_n^A} \) and \( \ket{\psi_n^B} \) are the normalized eigenvectors in the \( A \) and \( B \) subspaces.  

To analyze the topological properties, we perform a singular value decomposition (SVD) on the off-diagonal block \( h \) as $	h = U_A \Sigma U_B^{\dagger}$,   
where \( U_A \) and \( U_B \) are matrices of singular vectors, and \( \Sigma \) is a diagonal matrix of singular values. The columns of \( U_A \) and \( U_B \) form orthonormal bases for the subspaces associated with the eigenstates of the chiral symmetry operator.  

The quadrupole operator is defined as~\cite{Benalcazar_2022,pal2024}
\begin{equation}
	Q = \exp\left( -\frac{i 2\pi x y}{L_x L_y} \right)\ ,  
\end{equation}  
which measures the spatial distribution of charge in the system. The quadrupole moment is then defined within the sublattice sectors as  
\begin{equation}
	Q_{A,B} = \sum_{R, u \in A, B} \ket{R, u} Q \bra{R, u}.  
\end{equation}  
Using the singular eigenvectors from the SVD, the sublattice quadrupole moments are projected into the spaces of $U_{A,B}$ defined by:  
\begin{equation}
	\bar{Q}_{A,B} = U_{A,B}^{\dagger} Q  U_{A,B}\ .  
\end{equation}  

The quadrupolar winding number can be then computed as~\cite{Benalcazar_2022,pal2024}:  
\begin{equation}
	N_{xy} = \frac{1}{2\pi i} \operatorname{Tr} \log \left( \bar{Q}_{x,y}^A \bar{Q}_{x,y}^{B \dagger} \right)\ .  
\end{equation}  
This invariant captures the topological nature of the bulk and reflects the number of localized MCMs at the system's corners.  

\begin{figure}  
	\centering  
	\includegraphics[width=0.5\textwidth]{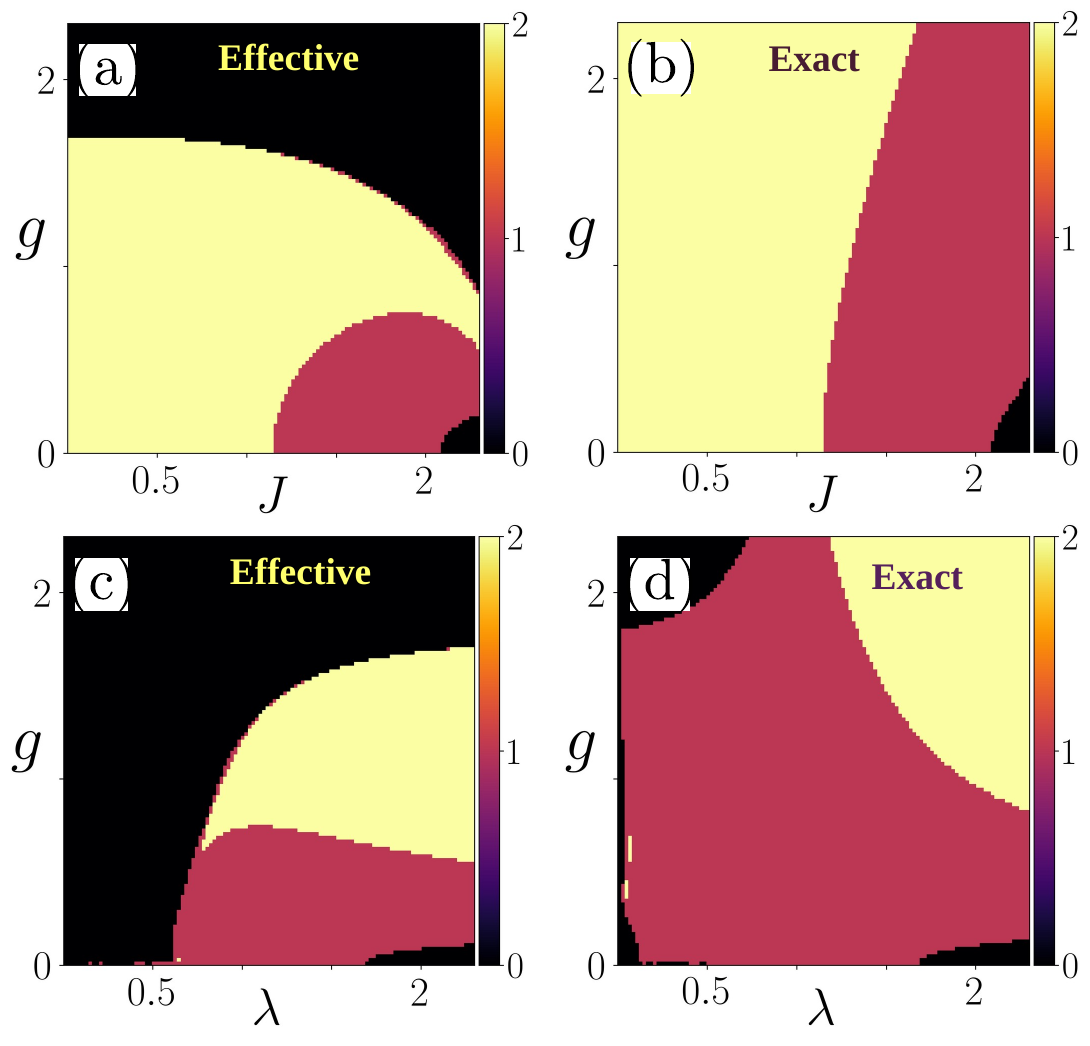}  
	\caption{Topological characterization of SOTSC phase via $N_{xy}$ is illustrated for both the exact and effective lattice regularized models. In panels (a) and (c), we depict $N_{xy}$ for the effective model in the $g$-$J$ plane choosing $\lambda = t$ and in the $g$-$\lambda$ plane at $J = 2t$, respectively. On the other hand, in panels (b) and (d), we display the corresponding results for the exact lattice model in the same parameter space.  In the phase diagrams, the yellow and pink regions correspond to \( N_{xy} = 2 \) and \( N_{xy} = 1 \) respectively. The phase diagram in panel (b) corroborates
	with the phase diagram of number of MCMs presented in Fig.~\ref{Fig2}(b). 
	The calculations are performed with chosen model parameters $\Delta_{0} = 0.5t$, $\epsilon_0 = t$, 
	and on a finite size system of $30 \times 30$ lattice sites.}  
	\label{Fig4}  
\end{figure}

We compute \( N_{xy} \) considering both the exact lattice model and the lattice-regularized effective model. In both the cases, \( N_{xy} \) achieves values 0, 1, and 2, corresponding to the yellow, pink, and black regions in Fig.~\ref{Fig4}, respectively. In Figs.~\ref{Fig4}(a) and \ref{Fig4}(c), we show the corresponding results of \( N_{xy} \) for the effective model in the \( g \)--\( J \) and \( g \)--\( \lambda \) planes, while Figs.~\ref{Fig4}(b) and \ref{Fig4}(d) represent the corresponding results for the exact lattice model. A clear correspondence emerges between the computed values of \( N_{xy} \) and the number of MCMs, consistent with the bulk-boundary correspondence as can be seen by comparing Fig.~\ref{Fig2}(b) and Fig.~\ref{Fig4}(b). In the second-order phase with eight-MCMs, we find \( N_{xy} = 2 \), while the four-MCM phase corresponds to \( N_{xy} = 1 \) and \( N_{xy} = 0 \) refers to the trivial gapped phase. Additionally, the exact model exhibits a broader topological region than the effective model. This mismatch arises because, in transitioning from the exact lattice model to the effective one, we simplify the system by retaining only the low-energy terms, specifically, those proportional to \( k_x \) and \( k_x^2 \), while neglecting higher-order contributions. As a result, although the two models qualitatively capture the same physics, their quantitative predictions differ. Here, $N_{xy}$ is employed within a square lattice geometry under PBC, where it remains well-defined and offers meaningful characterization of the topological phases under consideration~\cite{Benalcazar_2022}. In general, one can also calculate Bott index that provides a more robust characterization, particularly for systems with complex geometries~\cite{Bott_Luo_2025}.

\section{Low energy edge-theory}\label{Sec:V}
To further understand the topological phase transitions and tunability of the MCMs by system parameters, we derive the effective low-energy edge Hamiltonian from the effective low-energy bulk Hamiltonian as evaluated in Sec.~\ref{Sec:III}. In this section, we compute the edge Hamiltonian across the four edges of our 2D heterostructure, as schematically shown in Fig.~\ref{Fig1}, and analyze 
the corner mode solutions.

\subsection{Edge Hamiltonian}
In order to proceed for the edge theory, we partition the Hamiltonian Eq.~(\ref{eq:EffH1}) into two parts: the unperturbed part $H_0$ and the perturbed part $H_p$. For edge-I, we impose OBC along the $x$-direction and PBC along the $y$-direction. Hence, $k_x=-i\partial_x$. The unperturbed Hamiltonian is then given by~\cite{zhong2018,Ghosh_2021b}:
\begin{equation}
	H_0(-i\partial_x) = (m' - t\partial_x^2)\Gamma_1 - 2i\lambda_x \partial_x \Gamma_5\ ,
\end{equation}
where $m' = (\epsilon_0 - 4t)$, which is assumed to be negative to satisfy the Fu-Kane criterion~\cite{FuKane_2007}.

The perturbed Hamiltonian can be expressed as:
\begin{equation}
	\begin{aligned}
		&H_p(-i\partial_x, k_y, g_x, g_y, J, \Delta_0) 
		= \frac{t}{4}(g_x^2+g_y^2)\Gamma_1  -\frac{t}{2}(-ig_x\partial_x\\
		&+g_yk_y)\Gamma_7
		 + J\Gamma_3 + 2\lambda_yk_y\Gamma_6
      + \Delta_0 \big( -\partial_x^2\Gamma_2 + \frac{1}{2}[g_yk_y\\
      &+ig_x\partial_x]\Big)\Gamma_4
	    - \lambda_xg_x\Gamma_8 - \lambda_y g_y \Gamma_9 \ .
	\end{aligned}
\end{equation}
Here, we consider $g_x$, $g_y$, $J$, and $\Delta_0$ to be small parameters compared to $m^{\prime}, t, \lambda_{x}$. The components multiplied by these terms are also assumed to be small.

By solving the unperturbed part $H_0$, we obtain the eigenvectors and use them in the perturbative calculations for $H_p$. The resulting edge-I Hamiltonian is given as (see Appendix~\ref{Appendix-D} for details):
\begin{equation}
	H_{\text{edge}}^{\text{I}} = \frac{\Delta_0 m'}{t} \tau_x s_0 - \lambda_y g_y \tau_z s_0 - 2\lambda_y k_y \tau_z s_z - \frac{\Delta_0}{2} g_y k_y \tau_x s_z\ .
\end{equation}

The generalized edge Hamiltonian for edge $j$ (where $j = \text{I, II, III, IV}$) can be written as:
\begin{equation}
	H_{\text{edge}}^j = A_j \tau_x s_0 + B_j \tau_z s_0 + C_j \tau_z s_z + D_j \tau_x s_z + E_j \tau_0 s_x\ .
\end{equation}
where, the coefficients are given by:
\begin{align*}
	A_j &= \left\{\Delta_0 m'/t, -\Delta_0 m'/t, \Delta_0 m'/t, -\Delta_0 m'/t \right\}\ , \\
	B_j &= \left\{-\lambda_y g_y, \lambda_x g_x, -\lambda_y g_y, \lambda_x g_x \right\}\ , \\
	C_j &= \left\{-2\lambda_y k_y, 2\lambda_x k_x, -2\lambda_y k_y, 2\lambda_x k_x \right\}\ , \\
	D_j &= \left\{-\Delta_0 g_y k_y/2, \Delta_0 g_x k_x/2, -\Delta_0 g_y k_y/2, \Delta_0 g_x k_x/2 \right\}\ , \\
	E_j &= \left\{0, J, 0, J \right\}.
\end{align*}

To simplify the analysis, we assume a homogeneous and symmetric spin texture with \( g_x = g_y = g \) and equal SOC strengths of QSHI along both spatial directions, \( \lambda_x = \lambda_y = \lambda \), and set \( k_x = k_y = 0 \). Under these assumptions, edges I and III remain trivially gapped, with an energy spectrum given by \( E_{\text{I,III}} = \pm \sqrt{\Delta_0^2 {m'}^2 + g^2 \lambda^2 t^2}/t \), which does not allow for any gap closing. In contrast, the energy spectrum for edges II and IV is given by \( E_{\text{II,IV}} = \pm J \pm \sqrt{\Delta_0^2 {m'}^2 + g^2 \lambda^2 t^2}/t \), where a gap closing occurs when the condition \( J = \pm \sqrt{\Delta_0^2 {m'}^2 + g^2 \lambda^2 t^2}/t \) is satisfied. This condition marks a possible topological phase transition along edges II and IV.
Hence, the mass gap of edges II and IV can be tuned and as a result MCMs appear at the corners of the 2D domain due to Jackiw-Rebbi theorem~\cite{Jackiw_Rebbi_1976}. This gap closing behavior is depicted in Fig.~\ref{Fig3}(b) for a nanoribbon within a lattice-regularized model, which exhibits qualitative agreement with the low-energy edge theory. From Figs.~\ref{Fig2}(a) and (b), we examine 
that the result (values of $J_{c_1}$) nearly matches for the topological transition from 8 to 4 MCMs.

\subsection{Majorana Corner Mode Solution}
To determine the approximate form of MCMs solution, we assume $g \approx 0$ and consider two cases (see Appendix~\ref{Appendix-E} for details):

$\textbf{Case I:}$ When ($J > \frac{\Delta_0 m'}{t})$ $\rightarrow$ 
		$\psi_c \propto \exp \left(-\frac{\Delta_0 m'}{2\lambda_y t} y\right) \phi_c$ (along edge-I) and,
		$\psi_c \propto \exp \left(-\frac{(Jt + \Delta_0 m')}{2\lambda_x t} x\right) \phi_c$ (along edge-II) where $\phi_c = \{1, -i, -i, 1\}^T$. This signifies the presence of a single Majorana mode per corner \cite{Ghosh_2021,Ghosh_2021b} with localization length $\left[\frac{(Jt + \Delta_0 m')}{2\lambda_x t}\right]^{-1}$ and $\left[\frac{\Delta_0 m'}{2\lambda_y t}\right]^{-1}$ along $x$ and $y$ directions respectively.

$\textbf{Case II:}$ If ($J < \frac{\Delta_0 m'}{t}$) $\rightarrow$
		$\psi_c \propto \exp \left(-\frac{\Delta_0 m'}{2\lambda_y t} y\right) \phi_{c_1} + \exp \left(-\frac{\Delta_0 m'}{2\lambda_y t} y\right) \phi_{c_2}$ (along edge-I),
and $\psi_c \propto \exp \left(-\frac{\Delta_0 m' - Jt}{2\lambda_y t} x\right) \phi_{c_1} + \exp \left(-\frac{(Jt + \Delta_0 m')}{2\lambda_y t} x\right) \phi_{c_2}$ (along edge-II).	
	Here, $\phi_{c_1} = \{1, -i, -i, 1\}^T$ and $\phi_{c_2} = \{1, -i, i, -1\}^T$. In this scenario, the corner-mode solution is a superposition of both $\phi_{c_1}$ and $\phi_{c_2}$, corresponding to two Majorana modes per corner along each edge \cite{Ghosh_2021,Ghosh_2021b}.  In both these cases, we see that one can have different localization length of MCMs along different directions.

	\section{Summary and Conclusion}\label{Sec:VI}
	To summarize, in this article, we propose a theoretical framework for realizing a 2D SOTSC in a hybrid system comprising of a $d$-wave SC, a QSHI, and a noncollinear spin texture. This composite structure can host second-order topological superconducting phase with tunable MCMs due to the interplay of $d$-wave SC and spin texture. Crucially, the exchange strength of the spin texture acts as a tuning parameter for the MCMs, enabling control over their number at each corner. Depending on the system parameters, either one or two MCMs can appear per corner. The bulk topology of the system is characterized by the quadrupolar winding number $N_{xy}$, which takes a value of one when four MCMs are present and two when the system hosts eight MCMs. To gain deeper insight into the underlying physics, we perform a unitary transformation that leads to an effective low-energy Hamiltonian, revealing the emergence of an in-plane Zeeman field and effective SOC induced by the spin texture. This analytical result is qualitatively supported by the results obtained from the exact real-space numerical calculation, confirming the robustness of our theoretical model. For completeness, 
the effect of external Rashba SOC on the SOTSC phase is discussed in Appendix~\ref{Appendix-F}. The pairing mechanism responsible for the stability of SOTSC phase is further examined through an effective bulk pairing analysis, which highlights the interplay of emergent $s$-wave and $p$-wave pairing components. We further investigate a low energy edge theory from the effective model which captures the tunability of the MCMs and provides a qualitative understanding of the topological phase transitions.
	
  As far as practical feasibility of our setup is concerned, 
  magnetic adatoms (Mn/Cr) deposited on the surface of $s$-wave SC is experimentally shown to host first order MZMs \cite{Schneider2022}. Also, layer of magnetic adatoms(Fe/Cr/Nb) deposited on the surface of iron based superconducting substrate (such as FeSe, $\beta-\text{Fe}1.01$Se, LaFeAs, etc.) can host MZMs in high temperature platform \cite{Chatzopoulos2021,Medvedev2009,Kamihara2008,Peng2018}. Hence, our proposal of engineering SOTSC with tunable MCMs may be possible to test on a three layer system of Fe-based SC, magnetic adatoms (Mn/Cr) and QSHI (HgTe). However, we don't want to make any claim that our 2D tight binding lattice Hamiltonian mimicks such materials.

\vspace {-0.7cm}
	\subsection*{Acknowledgments}
	M.S. and A.S. acknowledge Department of Atomic Energy (DAE), Govt. of India for providing the financial support. M.S. and A.S. also acknowledge SAMKHYA: High-Performance Computing Facility provided by Institute of Physics, Bhubaneswar and the two workstations provided by the Institute of Physics, Bhubaneswar from the
	DAE APEX project for numerical computations. M.S. acknowledges A. Pal and P. Chatterjee for stimulating discussions regarding this work. 

\subsection*{Data Availibility Statement}
The datasets generated and analyzed during the current study are available from the corresponding author upon reasonable request.


\appendix	
	\section{Various limit of Effective Hamiltonian}
\begin{figure}
	\centering
	\subfigure{\includegraphics[width=0.5\textwidth]{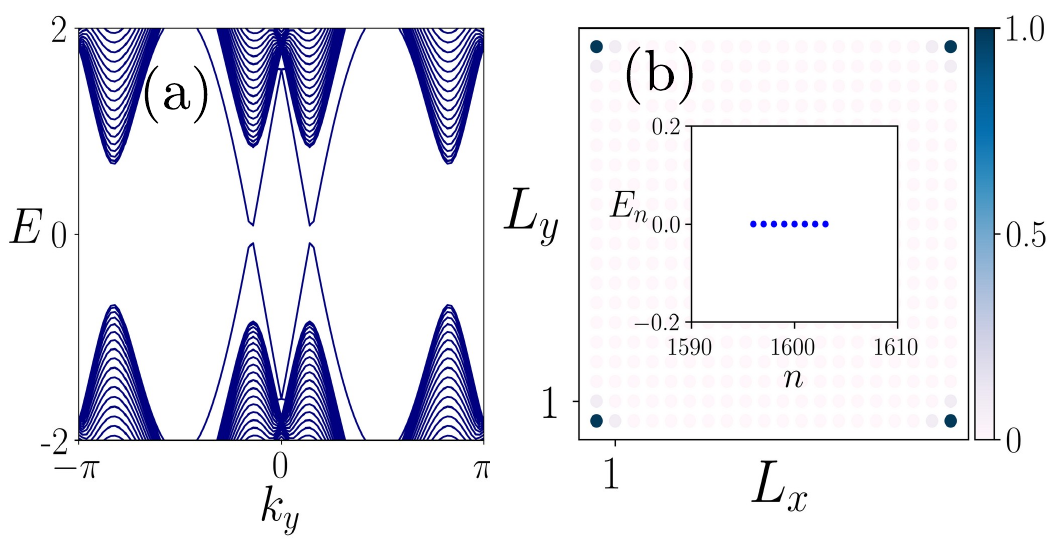}}
\caption{(a) We show the gapped edge spectrum of the lattice-regularized model employing OBC along the $x$-direction and PBC along the $y$-direction. The model parameters used are 
$\lambda_{x}=\lambda_{y}=\lambda = 2t$, $g = 0.8$, $J = 1.2t$, $\epsilon_0 = t$, $\Delta_0 = 0$, and $t = 1$. Panel (b) depicts the emergence of eight MCMs, with two modes localized at each corner, 
in a system comprising of a $d$-wave SC and a QSHI. Parameters are set as $\lambda = t$, $\epsilon_0 = t$, $J = g = 0$, and $\Delta_0 = 0.5t$. The inset displays the eight MCMs in the zoomed-in eigenvalue spectrum and the corresponding LDOS distribution displays their corner localization.}
\label{Fig5}
\end{figure}

{\label{Appendix-A}}
In this appendix, as introduced in Sec.~\ref{Sec:II}, we present the results corresponding to few limits of the exact lattice Hamiltonian Eq.~(\ref{eq:HamiltonianExact}). In order to do that, we employ 
the lattice regularized version of Eq.~(\ref{eq:EffH1}) with PBC along $y$ and OBC along $x$ direction and present the corresponding results in Fig.~\ref{Fig5}(a). 
Here, we only consider the interaction between QSHI and non-collinear spin texture and the superconducting order parameter $\Delta_0=0$. In this case, the TRS is broken and the helical gapless edge states that are present in the QSHI phase are gapped out and the system becomes a trivial insulator~\cite{science-Bernevig,science-zhang}. In Fig.~\ref{Fig5}(b), we consider the two layer system with
$d$-wave SC and QSHI employing OBC along both $x$ and $y$ directions in Eq.~(\ref{eq:EffH1}) of the main text. We observe that total 8 MCMs appear in this SOTSC phase \ie two MCMs are localized at each corner of the 2D domain. This is further illustrated by the LDOS spectra in Fig.~\ref{Fig5}(b) and 8 zero-energy MCMs are depicted in the inset. This result is consistent with Ref.~\cite{zhong2018}. 

		\section{Derivation of Low Energy Effective Hamiltonian}{\label{Appendix-B}}
	Here, we derive the effective low energy Hamiltonian introduced in mixed space representation in Eq.~(\ref{eq:Low energy H}). By substituting $k_{x,y}\rightarrow -i \nabla_{x,y}$ in Eq.~(\ref{eq:Low energy H}), the Hamiltonian takes the form:
	\begin{equation}
		H_{L}=\xi_k\Gamma_1-\Delta_d\Gamma_2
		-2i\lambda_x\nabla_x\Gamma_5-2i\lambda_y\nabla_y\Gamma_6+J\textbf{S(r).s}\ ,		
	\end{equation}
	where, $\xi_k=\left[\epsilon_0-4t-t \left(\nabla_x^2+\nabla_y^2\right)\right]$ and $\Delta_d=\Delta_0 \left(\nabla_x^2-\nabla_y^2\right)$. The effective Hamiltonian can obtained by a unitary transformation $H_{\text{eff}}=U^{\dagger}H_LU$ where $U=\tau_0\sigma_0e^{-i\frac{\phi}{2}s_z}$ and is given as 
	
	\begin{equation}
		\begin{aligned}
			&H_{\text{eff}} =  -t\sum_{r_i = x, y} \Bigg\{\left\{ -\frac{1}{4} \left(\frac{\partial \phi}{\partial r_i} \right)^2 \right\} \Gamma_1 - \frac{1}{2} \Bigg( \frac{i\partial \phi}{\partial r_i} \nabla_{r_i} \\
			 &+ i \nabla_{r_i} \frac{\partial \phi}{\partial r_i} \Bigg) \Gamma_7 \Bigg\}
			+ \xi_k  + J \Gamma_3 - 2i \lambda_y \nabla_y \Gamma_6 
			- 2i \lambda_x \nabla_x \Gamma_5\\  
			&+ \Delta_d \Gamma_2 - i \frac{\Delta_0}{2}\left( \frac{\partial \phi}{\partial y} \nabla_y - \frac{\partial \phi}{\partial x} \nabla_x \right) \Gamma_{10} -\lambda_x \frac{\partial  \phi}{\partial x}\Gamma_8 -\lambda_y \frac{\partial  \phi}{\partial y}\Gamma_9 .
		\end{aligned} 
		\label{eq:A2}
	\end{equation}
	
	Here, we assume the spin texture as $\phi_r=g_xx+g_yy$. Hence, substituting this in Eq.~(\ref{eq:A2}), our effective Hamiltonian reduces to:
\begin{equation}
	\begin{aligned}
		H_{\text{eff}} = & \xi_{k_\text{eff}}\Gamma_1 + \Delta_{d_\text{eff}} +J_{\text{ex}}\Gamma_3  
		+ 2\lambda_x k_x \Gamma_5 + 2\lambda_y k_y \Gamma_6 \\
		&- \frac{t}{2} \mathbf{g} \cdot \mathbf{k} \Gamma_7 
		- \lambda_x g_x \Gamma_8 - \lambda_y g_y \Gamma_9 \ ,  
	\end{aligned}
	\label{eq:EffH}
\end{equation}
	This is Eq.~(\ref{eq:EffH1}) of the main text. 

\section{Derivation of Effective pairing}\label{Appendix-C}
Here, we derive the effective pairing responsible for the appearance of SOTSC phase hosting MCMs. Starting from the effective model in Eq.~(\ref{eq:EffH1}), our Hamiltonian can be represented as:
\begin{equation}
	H_{\text{eff}} =
	\begin{pmatrix}
		\xi_k' + f' + J\sigma_0 s_x & \Delta_d' \sigma_0 s_z \\
		\Delta_d' \sigma_0 s_z & -\xi_k' - f' + J
	\end{pmatrix}\ ,
	\end{equation}
where, $f'= 2\lambda_x k_x \sigma_x s_z + 2\lambda_y k_y \sigma_y s_0 - \lambda_x g_x \sigma_x s_0 - \lambda_y g_y \sigma_y s_z - \frac{t}{2} ({\bf{g}} \cdot {\bf{k}}) \sigma_z s_z$, $\xi_k'= \left[(\epsilon_0 - 4t) - \frac{t}{4}(g_x^2 + g_y^2)\right] \sigma_z s_0$ and $\Delta_d'= \frac{\Delta_0}{2} (g_y k_y - g_x k_x)$.
Here, we have neglected the higher-order terms like $k_x^2$ and $k_y^2$.

We now perform a duality transformation on this Hamiltonian as~\cite{Chatterjee_2024a}
\begin{equation}
	H_D = U_d^{\dagger} H_{\text{eff}} U_d\ ,
	\end{equation}
where, $U_d$ denotes the unitary transformation as
\begin{equation}
	U_d = \bar{\tau} \sigma_0 s_0,
	\end{equation}
with
\begin{equation}
	\bar{\tau} = \frac{1}{\sqrt{2}} 
	\begin{pmatrix}
		1 & -1 \\
		1 & 1
	\end{pmatrix}\,.
	\end{equation}

Then, the transformed dual Hamiltonian becomes:
\begin{equation}
	H_D =
	\begin{pmatrix}
		\epsilon_D + J \sigma_0 s_x & \Delta_D \\
		\Delta_D & -\epsilon_D + J \sigma_0 s_x
	\end{pmatrix}\ ,
	\end{equation}
where,
\begin{align*}
	\epsilon_D &= \Delta_d' \sigma_0 s_z\ ,
\end{align*}
represents the dual kinetic energy. The off-diagonal term 
\begin{equation*}
	\Delta_D = -\xi_k' \sigma_z s_0 - f'
\end{equation*}
represents the effective pairing, which can be decomposed as
\begin{equation*}
	\Delta_D = \Delta_s + \Delta_p\ .
\end{equation*}

The terms $\Delta_s$ and $\Delta_p$ are given by:
\begin{align*}
	\Delta_s &= \left[(\epsilon_0 - 4t) - \frac{t}{2} g^2\right]\sigma_zs_0 - \lambda g (\sigma_x s_0 + \sigma_y s_z)\ , \\
	\Delta_p &= 2\lambda (k_x \sigma_x s_z + k_y \sigma_y s_0) - \frac{t}{2} g(k_x + k_y) \sigma_z s_z.
\end{align*}
where, we assume $\lambda_x=\lambda_y=\lambda$ as well as $g_x=g_y=g$. The term $\Delta_s$ denotes effective $s$-wave pairing. The first term in $\Delta_p$ denotes the effective  $p_x + i p_y$-type pairing generated by QSHI proportional to $\lambda$, while the second term, arising from the spin texture, represents a $p_x + p_y$-type pairing proportional to the pitch vector $g$. Such pairing stabilizes the SOTSC phase in the bulk. 

\section{Edge Theory}\label{Appendix-D}
Here, we show the contruction of the Hamiltonian for edge-II by conisdering PBC and OBC along $x$ and $y$ direction respectively. Rewriting Eq.~(\ref{eq:EffH1}) in terms of $H_0(-i\partial_y)$ and $H_p(-i\partial_y,k_x,g_x,g_y,J,\Delta_0)$ by neglecting $k_x^2$ and $k_y^2$ as:
\begin{equation}
	\begin{aligned}
	&H_0=[m'-t(\partial_y^2)]\Gamma_1-2i\lambda_y\partial_y \Gamma_6\ , \\
	&H_p=(t/4)(g_x^2+g_y^2)\Gamma_1-(t/2)(g_xk_x-ig_y\partial_y)\Gamma_7+
    \frac{\Delta_0}{2}[(\partial_y^2)\Gamma_2\\
    &-(ig_y\partial_y+g_xk_x)]\Gamma_{10}+J\Gamma_3-\lambda_xg_x \Gamma_8-\lambda_yg_y\Gamma_9+2\lambda_xk_x\Gamma_5\ ,
	\end{aligned}
\end{equation}

We solve $H_0$ exactly and treat $H_p$ as perturbation by assuming $g_x,g_y,\Delta_0,J$ to be small compared to the other energy scales. We consider $m'=\epsilon_0-4t<0$ to satisfy the Fu-Kane criteria \cite{FuKane_2007}. Assuming $\psi$ to be the zero energy eigen-state of $H_0$ and employing the boundary condition: $\psi(0)=\psi(\infty)=0$, we obtain:
\begin{equation}
	\psi_{\alpha}=Ce^{-\kappa_1y}\sin( \kappa_2 y)e^{ik_xx}\chi_{\alpha}\ ,
\end{equation}	
where $C=2\sqrt{\kappa_1(\kappa_1^2+\kappa_2^2)}/\kappa_2$ is the normalization factor,  $\kappa_1=\lambda_x/t$ and $\kappa_2=\sqrt{m^{'}/t-\kappa_1^2}$. $\chi_{\alpha}$ represents the eigenspinors given by:
\begin{equation}
	\chi_{1}=
	\begin{pmatrix}
		1\\0\\1\\0\\0\\0\\0\\0
	\end{pmatrix},
	\quad 
	\chi_{2}=
	\begin{pmatrix}
		0\\0\\0\\0\\1\\0\\1\\0
	\end{pmatrix},
	\quad 
	\chi_{3}=
	\begin{pmatrix}
		0\\1\\0\\1\\0\\0\\0\\0
	\end{pmatrix},
		\quad 
	\chi_{4}=
	\begin{pmatrix}
		0\\0\\0\\0\\0\\1\\0\\1
	\end{pmatrix}\ ,
\end{equation}
We obtain the edge Hamiltonian by operating these eigenspinors on the perturbed Hamiltonian as,
\begin{equation}
		H_{\text{edge}}^{\text{II}}=\int_{0}^{\infty} \psi_{\alpha}(y) H_p(-i\partial_y,k_x,g_x,g_y,J,\Delta_0) \psi_{\beta}(y) \,dy\ .
	\label{eq:D4}
\end{equation}
By evaluating Eq.~(\ref{eq:D4}), we obtain 
\begin{equation}
	\begin{aligned}
			H_{\text{edge}}^{\text{II}}=&J \tau_0\sigma_x-\frac{\Delta_0 m'}{t}\tau_x\sigma_0+\lambda_{x}g_{x}\tau_z\sigma_0\\
		&+2\lambda_xk_x
		\tau_z\sigma_z+(\Delta_0/2)g_xk_x\tau_x\sigma_z\ .
	\end{aligned} 
	\label{eq:D5}
\end{equation}

In the similar way one can derive the effective Hamiltonian for edge-I, where we consider PBC and OBC along $y$ and $x$-directions respectively. Proceeding the similar way as mentioned before and following the boundary condition: $\psi(0)=\psi(\infty)=0$, we obtain $\psi$ as:
\begin{equation}
	\psi_{\alpha}=Ce^{-\kappa_1x}\sin( \kappa_2 x)e^{ik_yy}\chi_{\alpha}\ ,
\end{equation}

and $\chi_{\alpha}$ is:
\begin{equation}
		\chi_{1}=
	\begin{pmatrix}
		i\\0\\1\\0\\0\\0\\0\\0
	\end{pmatrix},
	\quad 
	\chi_{2}=
	\begin{pmatrix}
		0\\0\\0\\0\\-i\\0\\1\\0
	\end{pmatrix},
	\quad 
	\chi_{3}=
	\begin{pmatrix}
		0\\-i\\0\\1\\0\\0\\0\\0
	\end{pmatrix},
	\quad 
	\chi_{4}=
	\begin{pmatrix}
		0\\0\\0\\0\\0\\i\\0\\1
	\end{pmatrix}\ ,
\end{equation}
The edge-I Hamiltonian can be written as:
\begin{align}
	H_{\text{edge}}^{\text{I}}=&\frac{\Delta_0 m'}{t}\tau_x\sigma_0-\lambda_{y}g_{y}\tau_z\sigma_0
	-2\lambda_yk_y
	\tau_z\sigma_z\nonumber\\
	&-(\Delta_0/2)g_yk_y\tau_x\sigma_z\ ,
\label{eq:D8}
\end{align}

Similarly, for edge-III and edge-IV, we proceed in the similar way as I and II but by considering boundary conditions $\psi(0)=\psi(-\infty)=0$ and thus we obtain:
\begin{equation}
	\begin{aligned}
			H_{\text{edge}}^{\text{III}}=&\frac{\Delta_0 m'}{t}\tau_x\sigma_0-\lambda_{y}g_{y}\tau_z\sigma_0
		-2\lambda_yk_y
		\tau_z\sigma_z\nonumber\\
		&-(\Delta_0/2)g_yk_y\tau_x\sigma_z\ ,\\
			H_{\text{edge}}^{\text{IV}}=&J \sigma_x-\frac{\Delta_0 m'}{t}\tau_x+\lambda_{x}g_{x}\tau_z
		+2\lambda_xk_x
		\tau_z\sigma_z\nonumber\\
		&+(\Delta_0/2)g_xk_x\tau_x\sigma_z\ .
	\end{aligned} 
\end{equation}

Afterwards, by choosing $g_x = g_y = g$, $\lambda_x = \lambda_y = \lambda$ and the momentum $k_x = k_y = 0 $, edges I and III remain trivially gapped, with an energy spectrum given by
\begin{equation}
E_{\text{I,III}} = \pm \frac{\sqrt{\Delta_0^2 {m'}^2 + g^2 \lambda^2 t^2}}{t}\ .
\end{equation}
Within this regime, the gap cannot close under any variation of system parameters. In contrast, the energy spectrum for edges II and IV takes the form
\begin{equation}
E_{\text{II,IV}} = \pm J \pm \frac{\sqrt{\Delta_0^2 {m'}^2 + g^2 \lambda^2 t^2}}{t}\ .
\end{equation}
Here, a gap closing occurs when the condition
\begin{equation}
J = \pm \frac{\sqrt{\Delta_0^2 {m'}^2 + g^2 \lambda^2 t^2}}{t}\ ,
\end{equation}
is satisfied suggesting possibility of opposite mass terms along two adjacent edges (\eg edges I and II). This gives rise to MCMs according to Jackiw-Rebbi theorem~\cite{Jackiw_Rebbi_1976}. 


\section{MCMs Solution}\label{Appendix-E}
In Appendix~\ref{Appendix-D}, we derive the edge Hamiltonian. Here we show the solution for the zero-energy MCMs. To obtain the corner state solution at the intersection of edge-I and edge-II, here we assume $g=g_x=g_y\approx0$ for simplicity and solve the corresponding edge Hamiltonian for zero-energy solution. We assume a solution of the form
\begin{equation}
	\psi_c \propto e^{-\kappa y}\phi_c \ ,
\end{equation} 	
Substituting $\psi_c$ in Eq.~(\ref{eq:D8}) for edge-I by considering $k_y=-i\partial_y$, we obtain:
\begin{equation}
	H_{\text{edge}}^{\text{I}}=\frac{\Delta_0 m'}{t}\tau_x\sigma_0-\lambda_yg_y\tau_z\sigma_0-2i\lambda_y\kappa \tau_z\sigma_z-\frac{\Delta_0g_yi\kappa}{2}\tau_x\sigma_z \ ,
\end{equation}
For $H_{\text{edge}}^{\text{I}}=0$,
we find the solutions as:
\begin{equation}
	\kappa=\left\{\frac{\Delta_0 m'}{2\lambda_y t}, \frac{\Delta_0 m'}{2\lambda_y t}, -\frac{\Delta_0 m'}{2\lambda_y t}, -\frac{\Delta_0 m'}{2\lambda_y t}\right\}\ .
\end{equation}
From the boundary condition as $\psi_c(\infty)=0$, we obtain two linearly independent solutions of the form:
$\phi_{c_1}=\left\{1,-i,-i,1\right\}^T$ and $\phi_{c_2}=\left\{1,-i,i,-1\right\}^T$. Hence, for edge-I, the corner mode solution takes the form:
\begin{equation}
	\psi_{C_1}=Ae^{\frac{-\Delta_0m'}{2\lambda_yt}y}\phi_{c_1}+Be^{\frac{-\Delta_0m'}{2\lambda_yt}y}\phi_{c_2}\ .
\end{equation}

Similarly, for edge-II, we substitute $k_x\rightarrow-i\partial_x$ and $\psi \propto e^{-\kappa x}\phi_c$ and proceed in the similar way to procure the 4 solutions of the form:
\begin{equation}
	\kappa=\left\{-\frac{\Delta_0 m'+Jt}{2\lambda_x t},\frac{\Delta_0 m'-Jt}{2\lambda_x t},\frac{Jt-\Delta_0 m'}{2\lambda_x t},\frac{\Delta_0 m'+Jt}{2\lambda_x t} \right\}\ ,
\end{equation}
Here, one can consider two situations:

\textbf{Case-I}:~$J>\Delta_0m^{'}/t$

Following the boundary condition $\psi_c(\infty)=0$, the spinors are:- $\phi_{c_1}=\left\{1,-i,-i,1\right\}^T$ and $\phi_{c_2}=\left\{1,i,-i,-1\right\}^T$. For edge-II, 
\begin{equation}
	\psi_{C_2}=Ce^{-\frac{Jt+\Delta_0m'}{2\lambda_xt}x}\phi_{c_1}+De^{-\frac{Jt-\Delta_0m'}{2\lambda_xt}x}\phi_{c_2}\ ,
\end{equation}

As both $\psi_{C_1}=\psi_{C_2}=0$ at $x=y=0$, so we obtain $B=0, D=0$ and $A=C$. Hence, the solution at the intersection becomes:
\begin{equation}
	\begin{aligned}
	&\psi_{C}\propto e^{-(Jt+\Delta_0 m')x/2\lambda t} \phi_c  \rightarrow  \text{along edge-II}\ ,\\
	&\psi_{C}\propto e^{-\Delta_0 m'y/2\lambda t} \phi_c  \rightarrow  \text{along edge-I}\ ,
	\end{aligned}
\end{equation}
where, $\left[{\Delta_0 m'/2\lambda t}\right]^{-1}$ and $\left[{(Jt+\Delta_0 m')/2\lambda t}\right]^{-1}$ are the localization lengths along $y$ and $x$ direction respectively. Similarly, at the other intersections also we can find the MCMs solution and it is evident that MCM decays along $x$ and $y$ directions separately and there is one MCM localized per corner.

\textbf{Case-II}: $J<\Delta_0m'/t$

Following the boundary conditions, one can obtain two linearly independent solutions as $\phi_{c_1}=\left\{1,-i,-i,1\right\}^T$ and $\phi_{c_2}=\left\{1,-i,i,-1\right\}^T$ and $\psi_{C_{2}}$ takes the form:
\begin{equation}
	\psi_{C_2}=Ce^{-\frac{Jt+\Delta_0m'}{2\lambda_xt}x}\phi_{c_1}+De^{-\frac{\Delta_0m'-Jt}{2\lambda_xt}x}\phi_{c_2}\ ,
\end{equation}.
At $x=y=0$, $\psi_{C_1}=\psi_{C_2}=0$. By satisfying this we obtain $A=C$ and $B=D$. 

Hence the MCM solution becomes:
\begin{equation}
	\begin{aligned}
		& \psi_{C} \propto A e^{-\frac{Jt + \Delta_0 m'}{2\lambda_x t} x}\phi_{c_1} + B e^{-\frac{\Delta_0 m'-Jt}{2\lambda_x t} x}\phi_{c_2}  \rightarrow  \text{along edge-II}\ , \\
		& \psi_{C} \propto A e^{-\frac{\Delta_0 m'}{2 \lambda t} y} \phi_{c_1} + B e^{-\frac{\Delta_0 m'}{2 \lambda t} y}\phi_{c_2}  \rightarrow  \text{along edge-I}\ .
	\end{aligned}
\end{equation}

Note that, in contrast to case-I, where we have only one MCM at the intersection, here we realize two modes per individual corner as along both $x$ and $y$ direction the MCMs exist as a linear superposition. The region $J>\Delta_0m'/t$ (one MCM per corner) and  $J<\Delta_0m'/t$ (2 MCMs per corner) match well with Fig.~\ref{Fig2}(b) of Sec.~\ref{Sec:II}.

\begin{figure}
	\centering
	\includegraphics[width=0.5\textwidth]{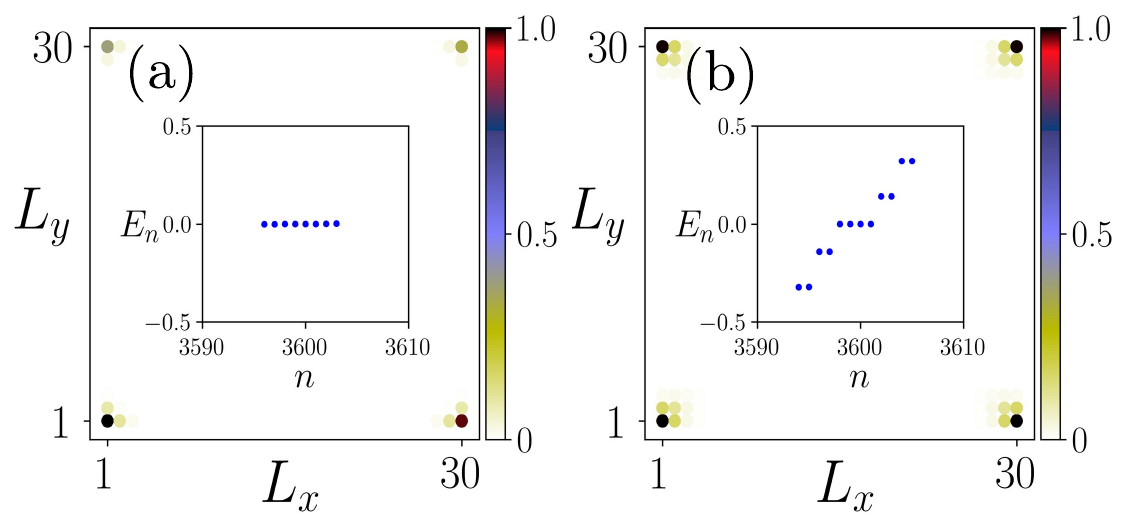}
	\caption{Panels (a) and (b) refer to regions I and II, containing 8 and 4 MCMs respectively, as highlighted in the insets. Rashba SOC is set to $\alpha_R = 0.05t$. In panel (a), the exchange 
	interaction is chosen as $J = 0.3t$, while in panel (b), $J = 1.4t$. All other model parameters are fixed at $g = 0.1$, $\Delta_0 = 0.5t$, and $\lambda_x = \lambda_y = \lambda = \epsilon_0 = t$. 
	We show the LDOS spectra considering a finite size $30 \times 30$ square lattice. The insets in both the panels display the zoomed-in eigenvalue spectra near zero energy.}
	\label{Fig6}
\end{figure}

\section{Effect of Rashba-SOC}\label{Appendix-F}
Here, we incorporate the Rashba SOC term separately in Eq.~(\ref{eq:HamiltonianExact}) of the main text and we study the behaviour of the MCMs in presence of this extra parameter. In practical point 
of view one can add a layer of material with Rashba SOC. After adding the Rashba SOC term, the Hamiltonian takes the form:
\begin{equation}
	\begin{aligned}
		H = \sum_{i,j} c_{i,j}^{\dagger} \bigg[ & \left(\epsilon_0 \Gamma_1+J(\Gamma_3\cos \phi_{i,j}+\Gamma_4 \sin \phi_{i,j})\right)c_{i,j} \\
		& -\left(t \Gamma_1+i\lambda_x\Gamma_5-\Delta_0\Gamma_2-i(\alpha_R/2)\Gamma_1^r \right)c_{i+1,j} \\
		& -\left(t\Gamma_1+i\lambda_y\Gamma_6+\Delta_0\Gamma_2+i(\alpha_R/2)\Gamma_2^r  \right)c_{i,j+1} \bigg] + h.c.\,
		\label{eq:Hamiltonian}
	\end{aligned}
\end{equation}
where, $\Gamma_1^r=\tau_z\sigma_0s_y$ and $\Gamma_2^r =\tau_z\sigma_0s_x$ denote the gamma matrices corresponding to the Rashba SOC in $y$ and $x$-directions respectively and $\alpha_R$ 
denotes the Rashba SOC strength. Other gamma matrices carry the same definitions as mentioned in the main text. By adding the Rashba term, we observe that our results remain qualitatively 
similar \ie the SOTSC phase anchoring 4 and 8 MCMs can still be seen with finite $\alpha_R$. 
In Fig.~\ref{Fig6}(a) and Fig.~\ref{Fig6}(b), we depict the number of MCMs (in the insets) and the corresponding LDOS distribution in the presence of Rashba SOC considering the region-I and 
region-II respectively as discussed in Sec.~\ref{Sec:II} of the main text. Note that, in Fig.~\ref{Fig6}(b), the LDOS spectra is slightly smeared out around the corners. Also, some bulk states appear 
within the zoomed-in eigenvalue spectra near zero energy (see the inset of Fig.~\ref{Fig6}(b)) in contrast to inset of Fig.~\ref{Fig6}(a) where only MCMs can be seen around zero energy. The reason
can be attributed to the different bulk gap sizes at $J = 0.3t$ and $J = 1.4t$ resulting in different localization lengths of the MCMs (see Fig.~\ref{Fig2}(a) for reference). 

{\section{Effect of conical spin texture}\label{Appendix-G}}
In this appendix, we consider conical spin texture to investigate the effect of an out-of-plane component in the magnetic texture. We generalize the exchange field of the magnetic impurities by introducing a full angular dependence. The modified Hamiltonian reads:
\begin{equation}
	\begin{aligned}
		H = \sum_{i,j} c_{i,j}^\dagger \Big[ \, & 
		 J \big( \Gamma_3 \cos \phi_{i,j} \sin \theta 
		+ \Gamma_4 \sin \phi_{i,j} \sin \theta 
		+ \Gamma_7 \cos \theta \big)  \\  &\epsilon_0 \Gamma_1 c_{i,j}
		 - \left( t \Gamma_1 + i \lambda_x \Gamma_5 - \Delta_0 \Gamma_2 \right) c_{i+1,j} \\
		& - \left( t \Gamma_1 + i \lambda_y \Gamma_6 + \Delta_0 \Gamma_2 \right) c_{i,j+1} \, \Big] 
		+ \text{h.c.}\ .
	\end{aligned}
	\label{eq:Hamiltonian_New_SSD}
\end{equation}
Here, the polar angle is defined as \(\phi_{i,j} = g_x x + g_y y\), and the Gamma matrices are: $\Gamma_1 = \tau_z \sigma_z s_0$, $\Gamma_2 = \tau_x \sigma_0 s_0$, $\Gamma_3 = \tau_0 \sigma_0 s_x$, $\Gamma_4 = \tau_0 \sigma_0 s_y$, $\Gamma_5 = \tau_z \sigma_x s_z$, $\Gamma_6 = \tau_z \sigma_y s_0$, $\Gamma_7 = \tau_0 \sigma_0 s_z$.

{Applying the unitary transformation \( U = \tau_0 \sigma_0 e^{-i\frac{\phi}{2}s_z} \), the Hamiltonian inbEq.~(\ref{eq:Hamiltonian_New_SSD}) becomes \( H_{\text{eff}} = U^\dagger H U \). Employing the lattice regularization approximations \( k_x^2 \approx 2(1 - \cos k_x) \) and \( k_x \approx \sin k_x \), we obtain the momentum-space Hamiltonian as
\begin{equation}
	\begin{aligned}
		H_{\text{lat}} = &\left[ \epsilon_0 - 2t \cos(k_x) - 2t \cos(k_y) \right] \Gamma_1 - \lambda_x g_x \Gamma_8 - \lambda_y g_y \Gamma_9  \\
		& - \frac{t}{2} (g_x \sin k_x + g_y \sin k_y) \Gamma_7 + J ( \sin \theta \Gamma_3 + \cos \theta \Gamma_7 ) \\
		& + 2 \lambda_x \sin k_x \Gamma_5 + 2 \lambda_y \sin k_y \Gamma_6 \\
		& + \Delta_0 \left[ (2\cos k_y - 2\cos k_x) \Gamma_2 + \frac{1}{2}(g_y \sin k_y - g_x \sin k_x) \Gamma_{10} \right]
	\end{aligned}
	\label{eq:H_latt_New}
\end{equation}
Here, the additional matrices are:
\(\Gamma_8 = \tau_z \sigma_x s_0\),
\(\Gamma_9 = \tau_z \sigma_y s_z\),
\(\Gamma_{10} = \tau_x \sigma_0 s_z\).

From Eq.~\eqref{eq:H_latt_New}, we observe the appearance of two Zeeman components: an in-plane part (\(J \sin \theta\)) and an out-of-plane part (\(J \cos \theta\)). We now investigate three representative cases:

\begin{itemize}
	\item \textbf{Case 1: \( \theta = 0 \)} \\
	
	The Zeeman term component points purely out-of-plane (along the \(z\)-axis). This configuration opens a gap in the energy eigenvalue 
	spectrum and eliminates the zero-energy modes that are present in the coplanar configuration. This is illustrated in Fig.~\ref{fig:Fig7}(a).
	
	\item \textbf{Case 2: \( \theta = \pi/2 \)} \\
	The Zeeman field lies entirely in-plane (along the \(x\)-axis). This scenario recovers the results discussed in the main text and 
	is shown again in Fig.~\ref{fig:Fig7}(b). Here, we observe a transition from eight zero-energy MCMs modes to four, and eventually to a trivial phase-indicating topological phase transitions.
	
	\item \textbf{Case 3: \( \theta = \pi/4 \)} \\
	The Zeeman field exhibits both in-plane and out-of-plane components with equal magnitude \(J/\sqrt{2}\). In this mixed configuration, zero-energy MCMs are initially gapped out, but reappear after a critical Zeeman strength \(J\) is reached, as shown in Fig.~\ref{fig:Fig7}(c). In this case, Fig.~\ref{fig:Fig7}(d) (inset) further confirms the existence of four such modes, localized at the corners of the 2D domain (see LDOS spectra in Fig.~\ref{fig:Fig7}(d)).
\end{itemize}

\begin{figure}
	\centering
	\includegraphics[scale=0.7]{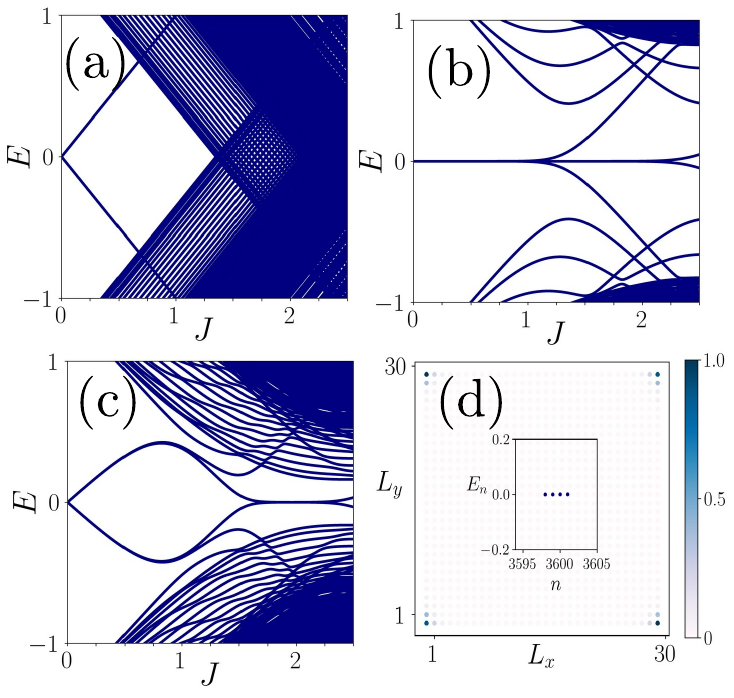}
	\caption{Eigenvalue spectrum is depicted as a function of \(J\) for (a) \(\theta = 0\), (b) \(\theta = \pi/2\), and (c) \(\theta = \pi/4\). In panel (d), we show the LDOS at \(J = 1.8t\), revealing four zero-energy modes localized at the corners of the 2D domain. We choose the other model parameters as \(\Delta_0 = 0.5t\), \(\epsilon_0 = t\), \(\lambda_x = \lambda_y = t\), \(g_x = g_y = g = t\), \(t = 1\) considering $30\times 30$ square lattice.}
	\label{fig:Fig7}
\end{figure}

{
{\section{Triangular Geometry}\label{Appendix-H}}}
In this appendix, in order to establish the robustness of the MCMs, we have computed the eigenvalue spectrum and LDOS in a different geometry that is triangular lattice. Our results show that the zero-energy modes remain well localized at the corners of the triangular lattice (see Fig.~\ref{fig:Fig8}(a) and Fig.~\ref{fig:Fig8}(b))~\cite{Chatterjee_2024b}. This confirms that the corner modes exhibit robustness even when the original crystalline symmetries are relaxed. The insets of Fig.~\ref{fig:Fig8}(a) and Fig.~\ref{fig:Fig8}(b) correspond to four and two zero-energy modes, respectively, which are localized at the corners of the triangular lattice. Hence, for other geometries as well, we find localized MCMs as the bulk topology is expected to be independent of the geometry.

\begin{figure}[H]
	\centering
	\includegraphics[scale=0.48]{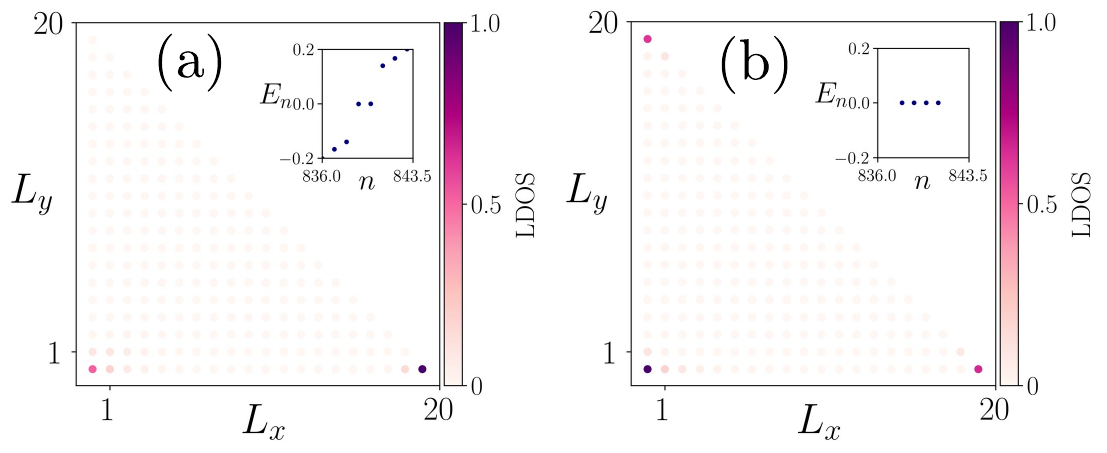}
	\caption{We illustrate the LDOS distribution where MCMs are localized at the corners of a triangular lattice. Panel (a) corresponds to the region-I ($J = 0.8t$) of the square lattice geometry discussed in the main text, while panel (b) denotes the region-II ($J = 1.8t$) of the same. The insets of panels (a) and (b) correspond to four and two zero-energy MCMs, respectively in the eigenvalue spectrum. The other model parameters are set as $\lambda_x = \lambda_y = t$, $\epsilon_0 = t$, $g_x = g_y = 0.2t$, $\Delta_0 = 0.5t$, and $t = 1$.}
	\label{fig:Fig8}
\end{figure}

\bibliography{bibfile}

\begin{thebibliography}{85}%
\makeatletter
\providecommand \@ifxundefined [1]{%
 \@ifx{#1\undefined}
}%
\providecommand \@ifnum [1]{%
 \ifnum #1\expandafter \@firstoftwo
 \else \expandafter \@secondoftwo
 \fi
}%
\providecommand \@ifx [1]{%
 \ifx #1\expandafter \@firstoftwo
 \else \expandafter \@secondoftwo
 \fi
}%
\providecommand \natexlab [1]{#1}%
\providecommand \enquote  [1]{``#1''}%
\providecommand \bibnamefont  [1]{#1}%
\providecommand \bibfnamefont [1]{#1}%
\providecommand \citenamefont [1]{#1}%
\providecommand \href@noop [0]{\@secondoftwo}%
\providecommand \href [0]{\begingroup \@sanitize@url \@href}%
\providecommand \@href[1]{\@@startlink{#1}\@@href}%
\providecommand \@@href[1]{\endgroup#1\@@endlink}%
\providecommand \@sanitize@url [0]{\catcode `\\12\catcode `\$12\catcode
  `\&12\catcode `\#12\catcode `\^12\catcode `\_12\catcode `\%12\relax}%
\providecommand \@@startlink[1]{}%
\providecommand \@@endlink[0]{}%
\providecommand \url  [0]{\begingroup\@sanitize@url \@url }%
\providecommand \@url [1]{\endgroup\@href {#1}{\urlprefix }}%
\providecommand \urlprefix  [0]{URL }%
\providecommand \Eprint [0]{\href }%
\providecommand \doibase [0]{http://dx.doi.org/}%
\providecommand \selectlanguage [0]{\@gobble}%
\providecommand \bibinfo  [0]{\@secondoftwo}%
\providecommand \bibfield  [0]{\@secondoftwo}%
\providecommand \translation [1]{[#1]}%
\providecommand \BibitemOpen [0]{}%
\providecommand \bibitemStop [0]{}%
\providecommand \bibitemNoStop [0]{.\EOS\space}%
\providecommand \EOS [0]{\spacefactor3000\relax}%
\providecommand \BibitemShut  [1]{\csname bibitem#1\endcsname}%
\let\auto@bib@innerbib\@empty
\bibitem [{\citenamefont {Kitaev}(2001)}]{Kitaev2001}%
  \BibitemOpen
  \bibfield  {author} {\bibinfo {author} {\bibfnamefont {A.~Y.}\ \bibnamefont
  {Kitaev}},\ }\bibfield  {title} {\enquote {\bibinfo {title} {Unpaired
  majorana fermions in quantum wires},}\ }\href {\doibase
  10.1070/1063-7869/44/10S/S29} {\bibfield  {journal} {\bibinfo  {journal}
  {Physics-Uspekhi}\ }\textbf {\bibinfo {volume} {44}},\ \bibinfo {pages} {131}
  (\bibinfo {year} {2001})}\BibitemShut {NoStop}%
\bibitem [{\citenamefont {Qi}\ and\ \citenamefont {Zhang}(2011)}]{Cheng_2011}%
  \BibitemOpen
  \bibfield  {author} {\bibinfo {author} {\bibfnamefont {X.-L.}\ \bibnamefont
  {Qi}}\ and\ \bibinfo {author} {\bibfnamefont {S.-C.}\ \bibnamefont {Zhang}},\
  }\bibfield  {title} {\enquote {\bibinfo {title} {Topological insulators and
  superconductors},}\ }\href {\doibase 10.1103/RevModPhys.83.1057} {\bibfield
  {journal} {\bibinfo  {journal} {Rev. Mod. Phys.}\ }\textbf {\bibinfo {volume}
  {83}},\ \bibinfo {pages} {1057--1110} (\bibinfo {year} {2011})}\BibitemShut
  {NoStop}%
\bibitem [{\citenamefont {Ivanov}(2001)}]{Ivanov2001}%
  \BibitemOpen
  \bibfield  {author} {\bibinfo {author} {\bibfnamefont {D.~A.}\ \bibnamefont
  {Ivanov}},\ }\bibfield  {title} {\enquote {\bibinfo {title} {Non-abelian
  statistics of half-quantum vortices in $\mathit{p}$-wave superconductors},}\
  }\href {\doibase 10.1103/PhysRevLett.86.268} {\bibfield  {journal} {\bibinfo
  {journal} {Phys. Rev. Lett.}\ }\textbf {\bibinfo {volume} {86}},\ \bibinfo
  {pages} {268--271} (\bibinfo {year} {2001})}\BibitemShut {NoStop}%
\bibitem [{\citenamefont {Nayak}\ \emph {et~al.}(2008)\citenamefont {Nayak},
  \citenamefont {Simon}, \citenamefont {Stern}, \citenamefont {Freedman},\ and\
  \citenamefont {Das~Sarma}}]{SDSarma2008}%
  \BibitemOpen
  \bibfield  {author} {\bibinfo {author} {\bibfnamefont {C.}~\bibnamefont
  {Nayak}}, \bibinfo {author} {\bibfnamefont {S.~H.}\ \bibnamefont {Simon}},
  \bibinfo {author} {\bibfnamefont {A.}~\bibnamefont {Stern}}, \bibinfo
  {author} {\bibfnamefont {M.}~\bibnamefont {Freedman}}, \ and\ \bibinfo
  {author} {\bibfnamefont {S.}~\bibnamefont {Das~Sarma}},\ }\bibfield  {title}
  {\enquote {\bibinfo {title} {Non-abelian anyons and topological quantum
  computation},}\ }\href {\doibase 10.1103/RevModPhys.80.1083} {\bibfield
  {journal} {\bibinfo  {journal} {Rev. Mod. Phys.}\ }\textbf {\bibinfo {volume}
  {80}},\ \bibinfo {pages} {1083--1159} (\bibinfo {year} {2008})}\BibitemShut
  {NoStop}%
\bibitem [{\citenamefont {Beenakker}(2013)}]{beenakker2013search}%
  \BibitemOpen
  \bibfield  {author} {\bibinfo {author} {\bibfnamefont {C.}~\bibnamefont
  {Beenakker}},\ }\bibfield  {title} {\enquote {\bibinfo {title} {Search for
  majorana fermions in superconductors},}\ }\href {\doibase
  10.1146/annurev-conmatphys-030212-184337} {\bibfield  {journal} {\bibinfo
  {journal} {Annual Review of Condensed Matter Physics}\ }\textbf {\bibinfo
  {volume} {4}},\ \bibinfo {pages} {113--136} (\bibinfo {year}
  {2013})}\BibitemShut {NoStop}%
\bibitem [{\citenamefont {Alicea}(2012)}]{Alicea_2012}%
  \BibitemOpen
  \bibfield  {author} {\bibinfo {author} {\bibfnamefont {J.}~\bibnamefont
  {Alicea}},\ }\bibfield  {title} {\enquote {\bibinfo {title} {New directions
  in the pursuit of majorana fermions in solid state systems},}\ }\href
  {\doibase 10.1088/0034-4885/75/7/076501} {\bibfield  {journal} {\bibinfo
  {journal} {Reports on Progress in Physics}\ }\textbf {\bibinfo {volume}
  {75}},\ \bibinfo {pages} {076501} (\bibinfo {year} {2012})}\BibitemShut
  {NoStop}%
\bibitem [{\citenamefont {Leijnse}\ and\ \citenamefont
  {Flensberg}(2012)}]{Leijnse_2012}%
  \BibitemOpen
  \bibfield  {author} {\bibinfo {author} {\bibfnamefont {M.}~\bibnamefont
  {Leijnse}}\ and\ \bibinfo {author} {\bibfnamefont {K.}~\bibnamefont
  {Flensberg}},\ }\bibfield  {title} {\enquote {\bibinfo {title} {Introduction
  to topological superconductivity and majorana fermions},}\ }\href {\doibase
  10.1088/0268-1242/27/12/124003} {\ \textbf {\bibinfo {volume} {27}},\
  \bibinfo {pages} {124003} (\bibinfo {year} {2012})}\BibitemShut {NoStop}%
\bibitem [{\citenamefont {Kitaev}(2009)}]{Kitaev2009}%
  \BibitemOpen
  \bibfield  {author} {\bibinfo {author} {\bibfnamefont {A.}~\bibnamefont
  {Kitaev}},\ }\bibfield  {title} {\enquote {\bibinfo {title} {{Periodic table
  for topological insulators and superconductors}},}\ }\href {\doibase
  10.1063/1.3149495} {\bibfield  {journal} {\bibinfo  {journal} {AIP Conference
  Proceedings}\ }\textbf {\bibinfo {volume} {1134}},\ \bibinfo {pages} {22--30}
  (\bibinfo {year} {2009})}\BibitemShut {NoStop}%
\bibitem [{\citenamefont {Geier}\ \emph {et~al.}(2018)\citenamefont {Geier},
  \citenamefont {Trifunovic}, \citenamefont {Hoskam},\ and\ \citenamefont
  {Brouwer}}]{Geier_2018}%
  \BibitemOpen
  \bibfield  {author} {\bibinfo {author} {\bibfnamefont {M.}~\bibnamefont
  {Geier}}, \bibinfo {author} {\bibfnamefont {L.}~\bibnamefont {Trifunovic}},
  \bibinfo {author} {\bibfnamefont {M.}~\bibnamefont {Hoskam}}, \ and\ \bibinfo
  {author} {\bibfnamefont {P.~W.}\ \bibnamefont {Brouwer}},\ }\bibfield
  {title} {\enquote {\bibinfo {title} {Second-order topological insulators and
  superconductors with an order-two crystalline symmetry},}\ }\href {\doibase
  10.1103/PhysRevB.97.205135} {\bibfield  {journal} {\bibinfo  {journal} {Phys.
  Rev. B}\ }\textbf {\bibinfo {volume} {97}},\ \bibinfo {pages} {205135}
  (\bibinfo {year} {2018})}\BibitemShut {NoStop}%
\bibitem [{\citenamefont {Song}\ \emph {et~al.}(2017)\citenamefont {Song},
  \citenamefont {Fang},\ and\ \citenamefont {Fang}}]{Zhong_2017}%
  \BibitemOpen
  \bibfield  {author} {\bibinfo {author} {\bibfnamefont {Z.}~\bibnamefont
  {Song}}, \bibinfo {author} {\bibfnamefont {Z.}~\bibnamefont {Fang}}, \ and\
  \bibinfo {author} {\bibfnamefont {C.}~\bibnamefont {Fang}},\ }\bibfield
  {title} {\enquote {\bibinfo {title} {$(d\ensuremath{-}2)$-dimensional edge
  states of rotation symmetry protected topological states},}\ }\href {\doibase
  10.1103/PhysRevLett.119.246402} {\bibfield  {journal} {\bibinfo  {journal}
  {Phys. Rev. Lett.}\ }\textbf {\bibinfo {volume} {119}},\ \bibinfo {pages}
  {246402} (\bibinfo {year} {2017})}\BibitemShut {NoStop}%
\bibitem [{\citenamefont {Khalaf}(2018)}]{Khalaf_2018}%
  \BibitemOpen
  \bibfield  {author} {\bibinfo {author} {\bibfnamefont {E.}~\bibnamefont
  {Khalaf}},\ }\bibfield  {title} {\enquote {\bibinfo {title} {Higher-order
  topological insulators and superconductors protected by inversion
  symmetry},}\ }\href {\doibase 10.1103/PhysRevB.97.205136} {\bibfield
  {journal} {\bibinfo  {journal} {Phys. Rev. B}\ }\textbf {\bibinfo {volume}
  {97}},\ \bibinfo {pages} {205136} (\bibinfo {year} {2018})}\BibitemShut
  {NoStop}%
\bibitem [{\citenamefont {Yan}\ \emph {et~al.}(2018)\citenamefont {Yan},
  \citenamefont {Song},\ and\ \citenamefont {Wang}}]{zhong2018}%
  \BibitemOpen
  \bibfield  {author} {\bibinfo {author} {\bibfnamefont {Z.}~\bibnamefont
  {Yan}}, \bibinfo {author} {\bibfnamefont {F.}~\bibnamefont {Song}}, \ and\
  \bibinfo {author} {\bibfnamefont {Z.}~\bibnamefont {Wang}},\ }\bibfield
  {title} {\enquote {\bibinfo {title} {Majorana corner modes in a
  high-temperature platform},}\ }\href {\doibase
  10.1103/PhysRevLett.121.096803} {\bibfield  {journal} {\bibinfo  {journal}
  {Phys. Rev. Lett.}\ }\textbf {\bibinfo {volume} {121}},\ \bibinfo {pages}
  {096803} (\bibinfo {year} {2018})}\BibitemShut {NoStop}%
\bibitem [{\citenamefont {Volovik}(2010)}]{Volovik2010}%
  \BibitemOpen
  \bibfield  {author} {\bibinfo {author} {\bibfnamefont {G.~E.}\ \bibnamefont
  {Volovik}},\ }\bibfield  {title} {\enquote {\bibinfo {title} {Topological
  superfluid \textsuperscript{3}he-b in magnetic field and ising variable},}\
  }\href {\doibase 10.1134/S0021364010040090} {\bibfield  {journal} {\bibinfo
  {journal} {JETP Letters}\ }\textbf {\bibinfo {volume} {91}},\ \bibinfo
  {pages} {201--205} (\bibinfo {year} {2010})}\BibitemShut {NoStop}%
\bibitem [{\citenamefont {Fu}\ and\ \citenamefont {Kane}(2008)}]{Fu2008}%
  \BibitemOpen
  \bibfield  {author} {\bibinfo {author} {\bibfnamefont {L.}~\bibnamefont
  {Fu}}\ and\ \bibinfo {author} {\bibfnamefont {C.~L.}\ \bibnamefont {Kane}},\
  }\bibfield  {title} {\enquote {\bibinfo {title} {Superconducting proximity
  effect and majorana fermions at the surface of a topological insulator},}\
  }\href {\doibase 10.1103/PhysRevLett.100.096407} {\bibfield  {journal}
  {\bibinfo  {journal} {Phys. Rev. Lett.}\ }\textbf {\bibinfo {volume} {100}},\
  \bibinfo {pages} {096407} (\bibinfo {year} {2008})}\BibitemShut {NoStop}%
\bibitem [{\citenamefont {Lutchyn}\ \emph {et~al.}(2010)\citenamefont
  {Lutchyn}, \citenamefont {Sau},\ and\ \citenamefont
  {Das~Sarma}}]{Lutchyn2010}%
  \BibitemOpen
  \bibfield  {author} {\bibinfo {author} {\bibfnamefont {R.~M.}\ \bibnamefont
  {Lutchyn}}, \bibinfo {author} {\bibfnamefont {J.~D.}\ \bibnamefont {Sau}}, \
  and\ \bibinfo {author} {\bibfnamefont {S.}~\bibnamefont {Das~Sarma}},\
  }\bibfield  {title} {\enquote {\bibinfo {title} {Majorana fermions and a
  topological phase transition in semiconductor-superconductor
  heterostructures},}\ }\href {\doibase 10.1103/PhysRevLett.105.077001}
  {\bibfield  {journal} {\bibinfo  {journal} {Phys. Rev. Lett.}\ }\textbf
  {\bibinfo {volume} {105}},\ \bibinfo {pages} {077001} (\bibinfo {year}
  {2010})}\BibitemShut {NoStop}%
\bibitem [{\citenamefont {Oreg}\ \emph {et~al.}(2010)\citenamefont {Oreg},
  \citenamefont {Refael},\ and\ \citenamefont {von Oppen}}]{Oreg2010}%
  \BibitemOpen
  \bibfield  {author} {\bibinfo {author} {\bibfnamefont {Y.}~\bibnamefont
  {Oreg}}, \bibinfo {author} {\bibfnamefont {G.}~\bibnamefont {Refael}}, \ and\
  \bibinfo {author} {\bibfnamefont {F.}~\bibnamefont {von Oppen}},\ }\bibfield
  {title} {\enquote {\bibinfo {title} {Helical liquids and majorana bound
  states in quantum wires},}\ }\href {\doibase 10.1103/PhysRevLett.105.177002}
  {\bibfield  {journal} {\bibinfo  {journal} {Phys. Rev. Lett.}\ }\textbf
  {\bibinfo {volume} {105}},\ \bibinfo {pages} {177002} (\bibinfo {year}
  {2010})}\BibitemShut {NoStop}%
\bibitem [{\citenamefont {Mourik}\ \emph {et~al.}(2012)\citenamefont {Mourik},
  \citenamefont {Zuo}, \citenamefont {Frolov}, \citenamefont {Plissard},
  \citenamefont {Bakkers},\ and\ \citenamefont {Kouwenhoven}}]{Mourik2012}%
  \BibitemOpen
  \bibfield  {author} {\bibinfo {author} {\bibfnamefont {V.}~\bibnamefont
  {Mourik}}, \bibinfo {author} {\bibfnamefont {K.}~\bibnamefont {Zuo}},
  \bibinfo {author} {\bibfnamefont {S.~M.}\ \bibnamefont {Frolov}}, \bibinfo
  {author} {\bibfnamefont {S.~R.}\ \bibnamefont {Plissard}}, \bibinfo {author}
  {\bibfnamefont {E.~P. A.~M.}\ \bibnamefont {Bakkers}}, \ and\ \bibinfo
  {author} {\bibfnamefont {L.~P.}\ \bibnamefont {Kouwenhoven}},\ }\bibfield
  {title} {\enquote {\bibinfo {title} {Signatures of majorana fermions in
  hybrid superconductor-semiconductor nanowire devices},}\ }\href {\doibase
  10.1126/science.1222360} {\bibfield  {journal} {\bibinfo  {journal}
  {Science}\ }\textbf {\bibinfo {volume} {336}},\ \bibinfo {pages} {1003--1007}
  (\bibinfo {year} {2012})}\BibitemShut {NoStop}%
\bibitem [{\citenamefont {Das}\ \emph {et~al.}(2012)\citenamefont {Das},
  \citenamefont {Ronen}, \citenamefont {Most}, \citenamefont {Oreg},
  \citenamefont {Heiblum},\ and\ \citenamefont {Shtrikman}}]{Das2012}%
  \BibitemOpen
  \bibfield  {author} {\bibinfo {author} {\bibfnamefont {A.}~\bibnamefont
  {Das}}, \bibinfo {author} {\bibfnamefont {Y.}~\bibnamefont {Ronen}}, \bibinfo
  {author} {\bibfnamefont {Y.}~\bibnamefont {Most}}, \bibinfo {author}
  {\bibfnamefont {Y.}~\bibnamefont {Oreg}}, \bibinfo {author} {\bibfnamefont
  {M.}~\bibnamefont {Heiblum}}, \ and\ \bibinfo {author} {\bibfnamefont
  {H.}~\bibnamefont {Shtrikman}},\ }\bibfield  {title} {\enquote {\bibinfo
  {title} {Zero-bias peaks and splitting in an al--inas nanowire topological
  superconductor as a signature of majorana fermions},}\ }\href {\doibase
  10.1038/nphys2479} {\bibfield  {journal} {\bibinfo  {journal} {Nature
  Physics}\ }\textbf {\bibinfo {volume} {8}},\ \bibinfo {pages} {887--895}
  (\bibinfo {year} {2012})}\BibitemShut {NoStop}%
\bibitem [{\citenamefont {Albrecht}\ \emph {et~al.}(2016)\citenamefont
  {Albrecht}, \citenamefont {Higginbotham}, \citenamefont {Madsen},
  \citenamefont {Kuemmeth}, \citenamefont {Jespersen}, \citenamefont
  {Nyg{\aa}rd}, \citenamefont {Krogstrup},\ and\ \citenamefont
  {Marcus}}]{Albrecht2016}%
  \BibitemOpen
  \bibfield  {author} {\bibinfo {author} {\bibfnamefont {S.~M.}\ \bibnamefont
  {Albrecht}}, \bibinfo {author} {\bibfnamefont {A.~P.}\ \bibnamefont
  {Higginbotham}}, \bibinfo {author} {\bibfnamefont {M.}~\bibnamefont
  {Madsen}}, \bibinfo {author} {\bibfnamefont {F.}~\bibnamefont {Kuemmeth}},
  \bibinfo {author} {\bibfnamefont {T.~S.}\ \bibnamefont {Jespersen}}, \bibinfo
  {author} {\bibfnamefont {J.}~\bibnamefont {Nyg{\aa}rd}}, \bibinfo {author}
  {\bibfnamefont {P.}~\bibnamefont {Krogstrup}}, \ and\ \bibinfo {author}
  {\bibfnamefont {C.~M.}\ \bibnamefont {Marcus}},\ }\bibfield  {title}
  {\enquote {\bibinfo {title} {Exponential protection of zero modes in majorana
  islands},}\ }\href {\doibase 10.1038/nature17162} {\bibfield  {journal}
  {\bibinfo  {journal} {Nature}\ }\textbf {\bibinfo {volume} {531}},\ \bibinfo
  {pages} {206--209} (\bibinfo {year} {2016})}\BibitemShut {NoStop}%
\bibitem [{\citenamefont {Pientka}\ \emph {et~al.}(2013)\citenamefont
  {Pientka}, \citenamefont {Glazman},\ and\ \citenamefont {von
  Oppen}}]{Felix2013}%
  \BibitemOpen
  \bibfield  {author} {\bibinfo {author} {\bibfnamefont {F.}~\bibnamefont
  {Pientka}}, \bibinfo {author} {\bibfnamefont {L.~I.}\ \bibnamefont
  {Glazman}}, \ and\ \bibinfo {author} {\bibfnamefont {F.}~\bibnamefont {von
  Oppen}},\ }\bibfield  {title} {\enquote {\bibinfo {title} {Topological
  superconducting phase in helical shiba chains},}\ }\href {\doibase
  10.1103/PhysRevB.88.155420} {\bibfield  {journal} {\bibinfo  {journal} {Phys.
  Rev. B}\ }\textbf {\bibinfo {volume} {88}},\ \bibinfo {pages} {155420}
  (\bibinfo {year} {2013})}\BibitemShut {NoStop}%
\bibitem [{\citenamefont {Nadj-Perge}\ \emph {et~al.}(2013)\citenamefont
  {Nadj-Perge}, \citenamefont {Drozdov}, \citenamefont {Bernevig},\ and\
  \citenamefont {Yazdani}}]{AliYazdani2013}%
  \BibitemOpen
  \bibfield  {author} {\bibinfo {author} {\bibfnamefont {S.}~\bibnamefont
  {Nadj-Perge}}, \bibinfo {author} {\bibfnamefont {I.~K.}\ \bibnamefont
  {Drozdov}}, \bibinfo {author} {\bibfnamefont {B.~A.}\ \bibnamefont
  {Bernevig}}, \ and\ \bibinfo {author} {\bibfnamefont {A.}~\bibnamefont
  {Yazdani}},\ }\bibfield  {title} {\enquote {\bibinfo {title} {Proposal for
  realizing majorana fermions in chains of magnetic atoms on a
  superconductor},}\ }\href {\doibase 10.1103/PhysRevB.88.020407} {\bibfield
  {journal} {\bibinfo  {journal} {Phys. Rev. B}\ }\textbf {\bibinfo {volume}
  {88}},\ \bibinfo {pages} {020407} (\bibinfo {year} {2013})}\BibitemShut
  {NoStop}%
\bibitem [{\citenamefont {Klinovaja}\ \emph {et~al.}(2013)\citenamefont
  {Klinovaja}, \citenamefont {Stano}, \citenamefont {Yazdani},\ and\
  \citenamefont {Loss}}]{DanielLoss2013}%
  \BibitemOpen
  \bibfield  {author} {\bibinfo {author} {\bibfnamefont {J.}~\bibnamefont
  {Klinovaja}}, \bibinfo {author} {\bibfnamefont {P.}~\bibnamefont {Stano}},
  \bibinfo {author} {\bibfnamefont {A.}~\bibnamefont {Yazdani}}, \ and\
  \bibinfo {author} {\bibfnamefont {D.}~\bibnamefont {Loss}},\ }\bibfield
  {title} {\enquote {\bibinfo {title} {Topological superconductivity and
  majorana fermions in rkky systems},}\ }\href {\doibase
  10.1103/PhysRevLett.111.186805} {\bibfield  {journal} {\bibinfo  {journal}
  {Phys. Rev. Lett.}\ }\textbf {\bibinfo {volume} {111}},\ \bibinfo {pages}
  {186805} (\bibinfo {year} {2013})}\BibitemShut {NoStop}%
\bibitem [{\citenamefont {Braunecker}\ and\ \citenamefont
  {Simon}(2013)}]{PascalSimon2013}%
  \BibitemOpen
  \bibfield  {author} {\bibinfo {author} {\bibfnamefont {B.}~\bibnamefont
  {Braunecker}}\ and\ \bibinfo {author} {\bibfnamefont {P.}~\bibnamefont
  {Simon}},\ }\bibfield  {title} {\enquote {\bibinfo {title} {Interplay between
  classical magnetic moments and superconductivity in quantum one-dimensional
  conductors: Toward a self-sustained topological majorana phase},}\ }\href
  {\doibase 10.1103/PhysRevLett.111.147202} {\bibfield  {journal} {\bibinfo
  {journal} {Phys. Rev. Lett.}\ }\textbf {\bibinfo {volume} {111}},\ \bibinfo
  {pages} {147202} (\bibinfo {year} {2013})}\BibitemShut {NoStop}%
\bibitem [{\citenamefont {Vazifeh}\ and\ \citenamefont
  {Franz}(2013)}]{MFranz2013}%
  \BibitemOpen
  \bibfield  {author} {\bibinfo {author} {\bibfnamefont {M.~M.}\ \bibnamefont
  {Vazifeh}}\ and\ \bibinfo {author} {\bibfnamefont {M.}~\bibnamefont
  {Franz}},\ }\bibfield  {title} {\enquote {\bibinfo {title} {Self-organized
  topological state with majorana fermions},}\ }\href {\doibase
  10.1103/PhysRevLett.111.206802} {\bibfield  {journal} {\bibinfo  {journal}
  {Phys. Rev. Lett.}\ }\textbf {\bibinfo {volume} {111}},\ \bibinfo {pages}
  {206802} (\bibinfo {year} {2013})}\BibitemShut {NoStop}%
\bibitem [{\citenamefont {Sau}\ and\ \citenamefont
  {Demler}(2013)}]{Eugene2013}%
  \BibitemOpen
  \bibfield  {author} {\bibinfo {author} {\bibfnamefont {J.~D.}\ \bibnamefont
  {Sau}}\ and\ \bibinfo {author} {\bibfnamefont {E.}~\bibnamefont {Demler}},\
  }\bibfield  {title} {\enquote {\bibinfo {title} {Bound states at impurities
  as a probe of topological superconductivity in nanowires},}\ }\href {\doibase
  10.1103/PhysRevB.88.205402} {\bibfield  {journal} {\bibinfo  {journal} {Phys.
  Rev. B}\ }\textbf {\bibinfo {volume} {88}},\ \bibinfo {pages} {205402}
  (\bibinfo {year} {2013})}\BibitemShut {NoStop}%
\bibitem [{\citenamefont {Pientka}\ \emph {et~al.}(2014)\citenamefont
  {Pientka}, \citenamefont {Glazman},\ and\ \citenamefont {von
  Oppen}}]{Felix2014}%
  \BibitemOpen
  \bibfield  {author} {\bibinfo {author} {\bibfnamefont {F.}~\bibnamefont
  {Pientka}}, \bibinfo {author} {\bibfnamefont {L.~I.}\ \bibnamefont
  {Glazman}}, \ and\ \bibinfo {author} {\bibfnamefont {F.}~\bibnamefont {von
  Oppen}},\ }\bibfield  {title} {\enquote {\bibinfo {title} {Unconventional
  topological phase transitions in helical shiba chains},}\ }\href {\doibase
  10.1103/PhysRevB.89.180505} {\bibfield  {journal} {\bibinfo  {journal} {Phys.
  Rev. B}\ }\textbf {\bibinfo {volume} {89}},\ \bibinfo {pages} {180505}
  (\bibinfo {year} {2014})}\BibitemShut {NoStop}%
\bibitem [{\citenamefont {P\"oyh\"onen}\ \emph {et~al.}(2014)\citenamefont
  {P\"oyh\"onen}, \citenamefont {Weststr\"om}, \citenamefont {R\"ontynen},\
  and\ \citenamefont {Ojanen}}]{TeemuOjanen2014}%
  \BibitemOpen
  \bibfield  {author} {\bibinfo {author} {\bibfnamefont {K.}~\bibnamefont
  {P\"oyh\"onen}}, \bibinfo {author} {\bibfnamefont {A.}~\bibnamefont
  {Weststr\"om}}, \bibinfo {author} {\bibfnamefont {J.}~\bibnamefont
  {R\"ontynen}}, \ and\ \bibinfo {author} {\bibfnamefont {T.}~\bibnamefont
  {Ojanen}},\ }\bibfield  {title} {\enquote {\bibinfo {title} {Majorana states
  in helical shiba chains and ladders},}\ }\href {\doibase
  10.1103/PhysRevB.89.115109} {\bibfield  {journal} {\bibinfo  {journal} {Phys.
  Rev. B}\ }\textbf {\bibinfo {volume} {89}},\ \bibinfo {pages} {115109}
  (\bibinfo {year} {2014})}\BibitemShut {NoStop}%
\bibitem [{\citenamefont {Reis}\ \emph {et~al.}(2014)\citenamefont {Reis},
  \citenamefont {Marchand},\ and\ \citenamefont {Franz}}]{MFranz2014}%
  \BibitemOpen
  \bibfield  {author} {\bibinfo {author} {\bibfnamefont {I.}~\bibnamefont
  {Reis}}, \bibinfo {author} {\bibfnamefont {D.~J.~J.}\ \bibnamefont
  {Marchand}}, \ and\ \bibinfo {author} {\bibfnamefont {M.}~\bibnamefont
  {Franz}},\ }\bibfield  {title} {\enquote {\bibinfo {title} {Self-organized
  topological state in a magnetic chain on the surface of a superconductor},}\
  }\href {\doibase 10.1103/PhysRevB.90.085124} {\bibfield  {journal} {\bibinfo
  {journal} {Phys. Rev. B}\ }\textbf {\bibinfo {volume} {90}},\ \bibinfo
  {pages} {085124} (\bibinfo {year} {2014})}\BibitemShut {NoStop}%
\bibitem [{\citenamefont {Hu}\ \emph {et~al.}(2015)\citenamefont {Hu},
  \citenamefont {Scalettar},\ and\ \citenamefont {Singh}}]{Rajiv2015}%
  \BibitemOpen
  \bibfield  {author} {\bibinfo {author} {\bibfnamefont {W.}~\bibnamefont
  {Hu}}, \bibinfo {author} {\bibfnamefont {R.~T.}\ \bibnamefont {Scalettar}}, \
  and\ \bibinfo {author} {\bibfnamefont {R.~R.~P.}\ \bibnamefont {Singh}},\
  }\bibfield  {title} {\enquote {\bibinfo {title} {Interplay of magnetic order,
  pairing, and phase separation in a one-dimensional spin-fermion model},}\
  }\href {\doibase 10.1103/PhysRevB.92.115133} {\bibfield  {journal} {\bibinfo
  {journal} {Phys. Rev. B}\ }\textbf {\bibinfo {volume} {92}},\ \bibinfo
  {pages} {115133} (\bibinfo {year} {2015})}\BibitemShut {NoStop}%
\bibitem [{\citenamefont {Hui}\ \emph {et~al.}(2015)\citenamefont {Hui},
  \citenamefont {Brydon}, \citenamefont {Sau}, \citenamefont {Tewari},\ and\
  \citenamefont {Sarma}}]{Sarma2015}%
  \BibitemOpen
  \bibfield  {author} {\bibinfo {author} {\bibfnamefont {H.-Y.}\ \bibnamefont
  {Hui}}, \bibinfo {author} {\bibfnamefont {P.~M.~R.}\ \bibnamefont {Brydon}},
  \bibinfo {author} {\bibfnamefont {J.~D.}\ \bibnamefont {Sau}}, \bibinfo
  {author} {\bibfnamefont {S.}~\bibnamefont {Tewari}}, \ and\ \bibinfo {author}
  {\bibfnamefont {S.~D.}\ \bibnamefont {Sarma}},\ }\bibfield  {title} {\enquote
  {\bibinfo {title} {Majorana fermions in ferromagnetic chains on the surface
  of bulk spin-orbit coupled s-wave superconductors},}\ }\href {\doibase
  10.1038/srep08880} {\bibfield  {journal} {\bibinfo  {journal} {Scientific
  Reports}\ }\textbf {\bibinfo {volume} {5}},\ \bibinfo {pages} {8880}
  (\bibinfo {year} {2015})}\BibitemShut {NoStop}%
\bibitem [{\citenamefont {Hoffman}\ \emph {et~al.}(2016)\citenamefont
  {Hoffman}, \citenamefont {Klinovaja},\ and\ \citenamefont
  {Loss}}]{Hoffman2016}%
  \BibitemOpen
  \bibfield  {author} {\bibinfo {author} {\bibfnamefont {S.}~\bibnamefont
  {Hoffman}}, \bibinfo {author} {\bibfnamefont {J.}~\bibnamefont {Klinovaja}},
  \ and\ \bibinfo {author} {\bibfnamefont {D.}~\bibnamefont {Loss}},\
  }\bibfield  {title} {\enquote {\bibinfo {title} {Topological phases of
  inhomogeneous superconductivity},}\ }\href {\doibase
  10.1103/PhysRevB.93.165418} {\bibfield  {journal} {\bibinfo  {journal} {Phys.
  Rev. B}\ }\textbf {\bibinfo {volume} {93}},\ \bibinfo {pages} {165418}
  (\bibinfo {year} {2016})}\BibitemShut {NoStop}%
\bibitem [{\citenamefont {Christensen}\ \emph {et~al.}(2016)\citenamefont
  {Christensen}, \citenamefont {Schecter}, \citenamefont {Flensberg},
  \citenamefont {Andersen},\ and\ \citenamefont {Paaske}}]{Jens2016}%
  \BibitemOpen
  \bibfield  {author} {\bibinfo {author} {\bibfnamefont {M.~H.}\ \bibnamefont
  {Christensen}}, \bibinfo {author} {\bibfnamefont {M.}~\bibnamefont
  {Schecter}}, \bibinfo {author} {\bibfnamefont {K.}~\bibnamefont {Flensberg}},
  \bibinfo {author} {\bibfnamefont {B.~M.}\ \bibnamefont {Andersen}}, \ and\
  \bibinfo {author} {\bibfnamefont {J.}~\bibnamefont {Paaske}},\ }\bibfield
  {title} {\enquote {\bibinfo {title} {Spiral magnetic order and topological
  superconductivity in a chain of magnetic adatoms on a two-dimensional
  superconductor},}\ }\href {\doibase 10.1103/PhysRevB.94.144509} {\bibfield
  {journal} {\bibinfo  {journal} {Phys. Rev. B}\ }\textbf {\bibinfo {volume}
  {94}},\ \bibinfo {pages} {144509} (\bibinfo {year} {2016})}\BibitemShut
  {NoStop}%
\bibitem [{\citenamefont {Sharma}\ and\ \citenamefont
  {Tewari}(2016)}]{Tewari2016}%
  \BibitemOpen
  \bibfield  {author} {\bibinfo {author} {\bibfnamefont {G.}~\bibnamefont
  {Sharma}}\ and\ \bibinfo {author} {\bibfnamefont {S.}~\bibnamefont
  {Tewari}},\ }\bibfield  {title} {\enquote {\bibinfo {title} {Yu-shiba-rusinov
  states and topological superconductivity in ising paired superconductors},}\
  }\href {\doibase 10.1103/PhysRevB.94.094515} {\bibfield  {journal} {\bibinfo
  {journal} {Phys. Rev. B}\ }\textbf {\bibinfo {volume} {94}},\ \bibinfo
  {pages} {094515} (\bibinfo {year} {2016})}\BibitemShut {NoStop}%
\bibitem [{\citenamefont {Andolina}\ and\ \citenamefont
  {Simon}(2017)}]{PascalSimon2017}%
  \BibitemOpen
  \bibfield  {author} {\bibinfo {author} {\bibfnamefont {G.~M.}\ \bibnamefont
  {Andolina}}\ and\ \bibinfo {author} {\bibfnamefont {P.}~\bibnamefont
  {Simon}},\ }\bibfield  {title} {\enquote {\bibinfo {title} {Topological
  properties of chains of magnetic impurities on a superconducting substrate:
  Interplay between the shiba band and ferromagnetic wire limits},}\ }\href
  {\doibase 10.1103/PhysRevB.96.235411} {\bibfield  {journal} {\bibinfo
  {journal} {Phys. Rev. B}\ }\textbf {\bibinfo {volume} {96}},\ \bibinfo
  {pages} {235411} (\bibinfo {year} {2017})}\BibitemShut {NoStop}%
\bibitem [{\citenamefont {Kaladzhyan}\ \emph {et~al.}(2017)\citenamefont
  {Kaladzhyan}, \citenamefont {Simon},\ and\ \citenamefont {Trif}}]{Simon2017}%
  \BibitemOpen
  \bibfield  {author} {\bibinfo {author} {\bibfnamefont {V.}~\bibnamefont
  {Kaladzhyan}}, \bibinfo {author} {\bibfnamefont {P.}~\bibnamefont {Simon}}, \
  and\ \bibinfo {author} {\bibfnamefont {M.}~\bibnamefont {Trif}},\ }\bibfield
  {title} {\enquote {\bibinfo {title} {Controlling topological
  superconductivity by magnetization dynamics},}\ }\href {\doibase
  10.1103/PhysRevB.96.020507} {\bibfield  {journal} {\bibinfo  {journal} {Phys.
  Rev. B}\ }\textbf {\bibinfo {volume} {96}},\ \bibinfo {pages} {020507}
  (\bibinfo {year} {2017})}\BibitemShut {NoStop}%
\bibitem [{\citenamefont {Theiler}\ \emph {et~al.}(2019)\citenamefont
  {Theiler}, \citenamefont {Bj\"ornson},\ and\ \citenamefont
  {Black-Schaffer}}]{Theiler2019}%
  \BibitemOpen
  \bibfield  {author} {\bibinfo {author} {\bibfnamefont {A.}~\bibnamefont
  {Theiler}}, \bibinfo {author} {\bibfnamefont {K.}~\bibnamefont {Bj\"ornson}},
  \ and\ \bibinfo {author} {\bibfnamefont {A.~M.}\ \bibnamefont
  {Black-Schaffer}},\ }\bibfield  {title} {\enquote {\bibinfo {title} {Majorana
  bound state localization and energy oscillations for magnetic impurity chains
  on conventional superconductors},}\ }\href {\doibase
  10.1103/PhysRevB.100.214504} {\bibfield  {journal} {\bibinfo  {journal}
  {Phys. Rev. B}\ }\textbf {\bibinfo {volume} {100}},\ \bibinfo {pages}
  {214504} (\bibinfo {year} {2019})}\BibitemShut {NoStop}%
\bibitem [{\citenamefont {Sticlet}\ and\ \citenamefont
  {Morari}(2019)}]{Cristian2019}%
  \BibitemOpen
  \bibfield  {author} {\bibinfo {author} {\bibfnamefont {D.}~\bibnamefont
  {Sticlet}}\ and\ \bibinfo {author} {\bibfnamefont {C.}~\bibnamefont
  {Morari}},\ }\bibfield  {title} {\enquote {\bibinfo {title} {Topological
  superconductivity from magnetic impurities on monolayer
  ${\mathrm{nbse}}_{2}$},}\ }\href {\doibase 10.1103/PhysRevB.100.075420}
  {\bibfield  {journal} {\bibinfo  {journal} {Phys. Rev. B}\ }\textbf {\bibinfo
  {volume} {100}},\ \bibinfo {pages} {075420} (\bibinfo {year}
  {2019})}\BibitemShut {NoStop}%
\bibitem [{\citenamefont {Mashkoori}\ and\ \citenamefont
  {Black-Schaffer}(2019)}]{Mashkoori2019}%
  \BibitemOpen
  \bibfield  {author} {\bibinfo {author} {\bibfnamefont {M.}~\bibnamefont
  {Mashkoori}}\ and\ \bibinfo {author} {\bibfnamefont {A.}~\bibnamefont
  {Black-Schaffer}},\ }\bibfield  {title} {\enquote {\bibinfo {title} {Majorana
  bound states in magnetic impurity chains: Effects of $d$-wave pairing},}\
  }\href {\doibase 10.1103/PhysRevB.99.024505} {\bibfield  {journal} {\bibinfo
  {journal} {Phys. Rev. B}\ }\textbf {\bibinfo {volume} {99}},\ \bibinfo
  {pages} {024505} (\bibinfo {year} {2019})}\BibitemShut {NoStop}%
\bibitem [{\citenamefont {M{\'e}nard}\ \emph {et~al.}(2019)\citenamefont
  {M{\'e}nard}, \citenamefont {Brun}, \citenamefont {Leriche}, \citenamefont
  {Trif}, \citenamefont {Debontridder}, \citenamefont {Demaille}, \citenamefont
  {Roditchev}, \citenamefont {Simon},\ and\ \citenamefont {Cren}}]{Menard2019}%
  \BibitemOpen
  \bibfield  {author} {\bibinfo {author} {\bibfnamefont {G.~C.}\ \bibnamefont
  {M{\'e}nard}}, \bibinfo {author} {\bibfnamefont {C.}~\bibnamefont {Brun}},
  \bibinfo {author} {\bibfnamefont {R.}~\bibnamefont {Leriche}}, \bibinfo
  {author} {\bibfnamefont {M.}~\bibnamefont {Trif}}, \bibinfo {author}
  {\bibfnamefont {F.}~\bibnamefont {Debontridder}}, \bibinfo {author}
  {\bibfnamefont {D.}~\bibnamefont {Demaille}}, \bibinfo {author}
  {\bibfnamefont {D.}~\bibnamefont {Roditchev}}, \bibinfo {author}
  {\bibfnamefont {P.}~\bibnamefont {Simon}}, \ and\ \bibinfo {author}
  {\bibfnamefont {T.}~\bibnamefont {Cren}},\ }\bibfield  {title} {\enquote
  {\bibinfo {title} {Yu-shiba-rusinov bound states versus topological edge
  states in {Pb/Si(111)}},}\ }\href {\doibase 10.1140/epjst/e2018-800056-3}
  {\bibfield  {journal} {\bibinfo  {journal} {The European Physical Journal
  Special Topics}\ }\textbf {\bibinfo {volume} {227}},\ \bibinfo {pages}
  {2303--2313} (\bibinfo {year} {2019})}\BibitemShut {NoStop}%
\bibitem [{\citenamefont {Mashkoori}\ \emph {et~al.}(2020)\citenamefont
  {Mashkoori}, \citenamefont {Pradhan}, \citenamefont {Bj\"ornson},
  \citenamefont {Fransson},\ and\ \citenamefont
  {Black-Schaffer}}]{Pradhan2020}%
  \BibitemOpen
  \bibfield  {author} {\bibinfo {author} {\bibfnamefont {M.}~\bibnamefont
  {Mashkoori}}, \bibinfo {author} {\bibfnamefont {S.}~\bibnamefont {Pradhan}},
  \bibinfo {author} {\bibfnamefont {K.}~\bibnamefont {Bj\"ornson}}, \bibinfo
  {author} {\bibfnamefont {J.}~\bibnamefont {Fransson}}, \ and\ \bibinfo
  {author} {\bibfnamefont {A.~M.}\ \bibnamefont {Black-Schaffer}},\ }\bibfield
  {title} {\enquote {\bibinfo {title} {Identification of topological
  superconductivity in magnetic impurity systems using bulk spin
  polarization},}\ }\href {\doibase 10.1103/PhysRevB.102.104501} {\bibfield
  {journal} {\bibinfo  {journal} {Phys. Rev. B}\ }\textbf {\bibinfo {volume}
  {102}},\ \bibinfo {pages} {104501} (\bibinfo {year} {2020})}\BibitemShut
  {NoStop}%
\bibitem [{\citenamefont {Teixeira}\ \emph {et~al.}(2020)\citenamefont
  {Teixeira}, \citenamefont {Kuzmanovski}, \citenamefont {Black-Schaffer},\
  and\ \citenamefont {da~Silva}}]{Teixeira2020}%
  \BibitemOpen
  \bibfield  {author} {\bibinfo {author} {\bibfnamefont {R.~L. R.~C.}\
  \bibnamefont {Teixeira}}, \bibinfo {author} {\bibfnamefont {D.}~\bibnamefont
  {Kuzmanovski}}, \bibinfo {author} {\bibfnamefont {A.~M.}\ \bibnamefont
  {Black-Schaffer}}, \ and\ \bibinfo {author} {\bibfnamefont {L.~G. G. V.~D.}\
  \bibnamefont {da~Silva}},\ }\bibfield  {title} {\enquote {\bibinfo {title}
  {Enhanced majorana bound states in magnetic chains on superconducting
  topological insulator edges},}\ }\href {\doibase 10.1103/PhysRevB.102.165312}
  {\bibfield  {journal} {\bibinfo  {journal} {Phys. Rev. B}\ }\textbf {\bibinfo
  {volume} {102}},\ \bibinfo {pages} {165312} (\bibinfo {year}
  {2020})}\BibitemShut {NoStop}%
\bibitem [{\citenamefont {Rex}\ \emph {et~al.}(2020)\citenamefont {Rex},
  \citenamefont {Gornyi},\ and\ \citenamefont {Mirlin}}]{Alexander2020}%
  \BibitemOpen
  \bibfield  {author} {\bibinfo {author} {\bibfnamefont {S.}~\bibnamefont
  {Rex}}, \bibinfo {author} {\bibfnamefont {I.~V.}\ \bibnamefont {Gornyi}}, \
  and\ \bibinfo {author} {\bibfnamefont {A.~D.}\ \bibnamefont {Mirlin}},\
  }\bibfield  {title} {\enquote {\bibinfo {title} {Majorana modes in
  emergent-wire phases of helical and cycloidal magnet-superconductor
  hybrids},}\ }\href {\doibase 10.1103/PhysRevB.102.224501} {\bibfield
  {journal} {\bibinfo  {journal} {Phys. Rev. B}\ }\textbf {\bibinfo {volume}
  {102}},\ \bibinfo {pages} {224501} (\bibinfo {year} {2020})}\BibitemShut
  {NoStop}%
\bibitem [{\citenamefont {Perrin}\ \emph {et~al.}(2021)\citenamefont {Perrin},
  \citenamefont {Civelli},\ and\ \citenamefont {Simon}}]{Perrin2021}%
  \BibitemOpen
  \bibfield  {author} {\bibinfo {author} {\bibfnamefont {V.}~\bibnamefont
  {Perrin}}, \bibinfo {author} {\bibfnamefont {M.}~\bibnamefont {Civelli}}, \
  and\ \bibinfo {author} {\bibfnamefont {P.}~\bibnamefont {Simon}},\ }\bibfield
   {title} {\enquote {\bibinfo {title} {Identifying majorana bound states by
  tunneling shot-noise tomography},}\ }\href {\doibase
  10.1103/PhysRevB.104.L121406} {\bibfield  {journal} {\bibinfo  {journal}
  {Phys. Rev. B}\ }\textbf {\bibinfo {volume} {104}},\ \bibinfo {pages}
  {L121406} (\bibinfo {year} {2021})}\BibitemShut {NoStop}%
\bibitem [{\citenamefont {Kobia\l{}ka}\ \emph {et~al.}(2020)\citenamefont
  {Kobia\l{}ka}, \citenamefont {Sedlmayr}, \citenamefont
  {Ma\ifmmode~\acute{s}\else \'{s}\fi{}ka},\ and\ \citenamefont
  {Doma\ifmmode~\acute{n}\else \'{n}\fi{}ski}}]{Nicholas2020}%
  \BibitemOpen
  \bibfield  {author} {\bibinfo {author} {\bibfnamefont {A.}~\bibnamefont
  {Kobia\l{}ka}}, \bibinfo {author} {\bibfnamefont {N.}~\bibnamefont
  {Sedlmayr}}, \bibinfo {author} {\bibfnamefont {M.~M.}\ \bibnamefont
  {Ma\ifmmode~\acute{s}\else \'{s}\fi{}ka}}, \ and\ \bibinfo {author}
  {\bibfnamefont {T.}~\bibnamefont {Doma\ifmmode~\acute{n}\else
  \'{n}\fi{}ski}},\ }\bibfield  {title} {\enquote {\bibinfo {title}
  {Dimerization-induced topological superconductivity in a rashba nanowire},}\
  }\href {\doibase 10.1103/PhysRevB.101.085402} {\bibfield  {journal} {\bibinfo
   {journal} {Phys. Rev. B}\ }\textbf {\bibinfo {volume} {101}},\ \bibinfo
  {pages} {085402} (\bibinfo {year} {2020})}\BibitemShut {NoStop}%
\bibitem [{\citenamefont {Chatterjee}\ \emph
  {et~al.}(2024{\natexlab{a}})\citenamefont {Chatterjee}, \citenamefont
  {Banik}, \citenamefont {Bera}, \citenamefont {Ghosh}, \citenamefont
  {Pradhan}, \citenamefont {Saha},\ and\ \citenamefont
  {Nandy}}]{Chatterjee_2024a}%
  \BibitemOpen
  \bibfield  {author} {\bibinfo {author} {\bibfnamefont {P.}~\bibnamefont
  {Chatterjee}}, \bibinfo {author} {\bibfnamefont {S.}~\bibnamefont {Banik}},
  \bibinfo {author} {\bibfnamefont {S.}~\bibnamefont {Bera}}, \bibinfo {author}
  {\bibfnamefont {A.~K.}\ \bibnamefont {Ghosh}}, \bibinfo {author}
  {\bibfnamefont {S.}~\bibnamefont {Pradhan}}, \bibinfo {author} {\bibfnamefont
  {A.}~\bibnamefont {Saha}}, \ and\ \bibinfo {author} {\bibfnamefont {A.~K.}\
  \bibnamefont {Nandy}},\ }\bibfield  {title} {\enquote {\bibinfo {title}
  {Topological superconductivity by engineering noncollinear magnetism in
  magnet/superconductor heterostructures: A realistic prescription for the
  two-dimensional kitaev model},}\ }\href {\doibase
  10.1103/PhysRevB.109.L121301} {\bibfield  {journal} {\bibinfo  {journal}
  {Phys. Rev. B}\ }\textbf {\bibinfo {volume} {109}},\ \bibinfo {pages}
  {L121301} (\bibinfo {year} {2024}{\natexlab{a}})}\BibitemShut {NoStop}%
\bibitem [{\citenamefont {Chatterjee}\ \emph
  {et~al.}(2024{\natexlab{b}})\citenamefont {Chatterjee}, \citenamefont
  {Ghosh}, \citenamefont {Nandy},\ and\ \citenamefont
  {Saha}}]{Chatterjee_2024b}%
  \BibitemOpen
  \bibfield  {author} {\bibinfo {author} {\bibfnamefont {P.}~\bibnamefont
  {Chatterjee}}, \bibinfo {author} {\bibfnamefont {A.~K.}\ \bibnamefont
  {Ghosh}}, \bibinfo {author} {\bibfnamefont {A.~K.}\ \bibnamefont {Nandy}}, \
  and\ \bibinfo {author} {\bibfnamefont {A.}~\bibnamefont {Saha}},\ }\bibfield
  {title} {\enquote {\bibinfo {title} {Second-order topological superconductor
  via noncollinear magnetic texture},}\ }\href {\doibase
  10.1103/PhysRevB.109.L041409} {\bibfield  {journal} {\bibinfo  {journal}
  {Phys. Rev. B}\ }\textbf {\bibinfo {volume} {109}},\ \bibinfo {pages}
  {L041409} (\bibinfo {year} {2024}{\natexlab{b}})}\BibitemShut {NoStop}%
\bibitem [{\citenamefont {Mondal}\ \emph {et~al.}(2023)\citenamefont {Mondal},
  \citenamefont {Ghosh}, \citenamefont {Nag},\ and\ \citenamefont
  {Saha}}]{Mondal2023}%
  \BibitemOpen
  \bibfield  {author} {\bibinfo {author} {\bibfnamefont {D.}~\bibnamefont
  {Mondal}}, \bibinfo {author} {\bibfnamefont {A.~K.}\ \bibnamefont {Ghosh}},
  \bibinfo {author} {\bibfnamefont {T.}~\bibnamefont {Nag}}, \ and\ \bibinfo
  {author} {\bibfnamefont {A.}~\bibnamefont {Saha}},\ }\bibfield  {title}
  {\enquote {\bibinfo {title} {Engineering anomalous floquet majorana modes and
  their time evolution in a helical shiba chain},}\ }\href {\doibase
  10.1103/PhysRevB.108.L081403} {\bibfield  {journal} {\bibinfo  {journal}
  {Phys. Rev. B}\ }\textbf {\bibinfo {volume} {108}},\ \bibinfo {pages}
  {L081403} (\bibinfo {year} {2023})}\BibitemShut {NoStop}%
\bibitem [{\citenamefont {Yang}\ \emph {et~al.}(2016)\citenamefont {Yang},
  \citenamefont {Stano}, \citenamefont {Klinovaja},\ and\ \citenamefont
  {Loss}}]{Jelena2016}%
  \BibitemOpen
  \bibfield  {author} {\bibinfo {author} {\bibfnamefont {G.}~\bibnamefont
  {Yang}}, \bibinfo {author} {\bibfnamefont {P.}~\bibnamefont {Stano}},
  \bibinfo {author} {\bibfnamefont {J.}~\bibnamefont {Klinovaja}}, \ and\
  \bibinfo {author} {\bibfnamefont {D.}~\bibnamefont {Loss}},\ }\bibfield
  {title} {\enquote {\bibinfo {title} {Majorana bound states in magnetic
  skyrmions},}\ }\href {\doibase 10.1103/PhysRevB.93.224505} {\bibfield
  {journal} {\bibinfo  {journal} {Phys. Rev. B}\ }\textbf {\bibinfo {volume}
  {93}},\ \bibinfo {pages} {224505} (\bibinfo {year} {2016})}\BibitemShut
  {NoStop}%
\bibitem [{\citenamefont {P\"oyh\"onen}\ \emph {et~al.}(2016)\citenamefont
  {P\"oyh\"onen}, \citenamefont {Weststr\"om}, \citenamefont {Pershoguba},
  \citenamefont {Ojanen},\ and\ \citenamefont {Balatsky}}]{Balatsky22016}%
  \BibitemOpen
  \bibfield  {author} {\bibinfo {author} {\bibfnamefont {K.}~\bibnamefont
  {P\"oyh\"onen}}, \bibinfo {author} {\bibfnamefont {A.}~\bibnamefont
  {Weststr\"om}}, \bibinfo {author} {\bibfnamefont {S.~S.}\ \bibnamefont
  {Pershoguba}}, \bibinfo {author} {\bibfnamefont {T.}~\bibnamefont {Ojanen}},
  \ and\ \bibinfo {author} {\bibfnamefont {A.~V.}\ \bibnamefont {Balatsky}},\
  }\bibfield  {title} {\enquote {\bibinfo {title} {Skyrmion-induced bound
  states in a $p$-wave superconductor},}\ }\href {\doibase
  10.1103/PhysRevB.94.214509} {\bibfield  {journal} {\bibinfo  {journal} {Phys.
  Rev. B}\ }\textbf {\bibinfo {volume} {94}},\ \bibinfo {pages} {214509}
  (\bibinfo {year} {2016})}\BibitemShut {NoStop}%
\bibitem [{\citenamefont {Subhadarshini}\ \emph {et~al.}()\citenamefont
  {Subhadarshini}, \citenamefont {Pal}, \citenamefont {Chatterjee},\ and\
  \citenamefont {Saha}}]{subhadarshini2025}%
  \BibitemOpen
  \bibfield  {author} {\bibinfo {author} {\bibfnamefont {M.}~\bibnamefont
  {Subhadarshini}}, \bibinfo {author} {\bibfnamefont {A.}~\bibnamefont {Pal}},
  \bibinfo {author} {\bibfnamefont {P.}~\bibnamefont {Chatterjee}}, \ and\
  \bibinfo {author} {\bibfnamefont {A.}~\bibnamefont {Saha}},\ }\bibfield
  {title} {\enquote {\bibinfo {title} {Identifying majorana edge and end modes
  in josephson junction of $p$-wave superconductor with magnetic barrier},}\
  }\href {https://arxiv.org/abs/2503.18362} {\ }\Eprint
  {http://arxiv.org/abs/2503.18362} {arXiv:2503.18362 [cond-mat.supr-con]}
  \BibitemShut {NoStop}%
\bibitem [{\citenamefont {Rachel}\ and\ \citenamefont
  {Wiesendanger}(2025)}]{RACHEL20251}%
  \BibitemOpen
  \bibfield  {author} {\bibinfo {author} {\bibfnamefont {S.}~\bibnamefont
  {Rachel}}\ and\ \bibinfo {author} {\bibfnamefont {R.}~\bibnamefont
  {Wiesendanger}},\ }\bibfield  {title} {\enquote {\bibinfo {title} {Majorana
  quasiparticles in atomic spin chains on superconductors},}\ }\href {\doibase
  https://doi.org/10.1016/j.physrep.2024.10.005} {\bibfield  {journal}
  {\bibinfo  {journal} {Physics Reports}\ }\textbf {\bibinfo {volume} {1099}},\
  \bibinfo {pages} {1--28} (\bibinfo {year} {2025})}\BibitemShut {NoStop}%
\bibitem [{\citenamefont {Lo~Conte}\ \emph {et~al.}(2024)\citenamefont
  {Lo~Conte}, \citenamefont {Wiebe}, \citenamefont {Rachel}, \citenamefont
  {Morr},\ and\ \citenamefont {Wiesendanger}}]{LoConte2024}%
  \BibitemOpen
  \bibfield  {author} {\bibinfo {author} {\bibfnamefont {R.}~\bibnamefont
  {Lo~Conte}}, \bibinfo {author} {\bibfnamefont {J.}~\bibnamefont {Wiebe}},
  \bibinfo {author} {\bibfnamefont {S.}~\bibnamefont {Rachel}}, \bibinfo
  {author} {\bibfnamefont {D.~K.}\ \bibnamefont {Morr}}, \ and\ \bibinfo
  {author} {\bibfnamefont {R.}~\bibnamefont {Wiesendanger}},\ }\bibfield
  {title} {\enquote {\bibinfo {title} {Magnet-superconductor hybrid quantum
  systems: a materials platform for topological superconductivity},}\ }\href
  {\doibase 10.1007/s40766-024-00060-1} {\bibfield  {journal} {\bibinfo
  {journal} {La Rivista del Nuovo Cimento}\ }\textbf {\bibinfo {volume} {47}},\
  \bibinfo {pages} {453--554} (\bibinfo {year} {2024})}\BibitemShut {NoStop}%
\bibitem [{\citenamefont {Yazdani}\ \emph {et~al.}(1997)\citenamefont
  {Yazdani}, \citenamefont {Jones}, \citenamefont {Lutz}, \citenamefont
  {Crommie},\ and\ \citenamefont {Eigler}}]{Eigler1997}%
  \BibitemOpen
  \bibfield  {author} {\bibinfo {author} {\bibfnamefont {A.}~\bibnamefont
  {Yazdani}}, \bibinfo {author} {\bibfnamefont {B.~A.}\ \bibnamefont {Jones}},
  \bibinfo {author} {\bibfnamefont {C.~P.}\ \bibnamefont {Lutz}}, \bibinfo
  {author} {\bibfnamefont {M.~F.}\ \bibnamefont {Crommie}}, \ and\ \bibinfo
  {author} {\bibfnamefont {D.~M.}\ \bibnamefont {Eigler}},\ }\bibfield  {title}
  {\enquote {\bibinfo {title} {Probing the local effects of magnetic impurities
  on superconductivity},}\ }\href {\doibase 10.1126/science.275.5307.1767}
  {\bibfield  {journal} {\bibinfo  {journal} {Science}\ }\textbf {\bibinfo
  {volume} {275}},\ \bibinfo {pages} {1767--1770} (\bibinfo {year}
  {1997})}\BibitemShut {NoStop}%
\bibitem [{\citenamefont {Yazdani}\ \emph {et~al.}(1999)\citenamefont
  {Yazdani}, \citenamefont {Howald}, \citenamefont {Lutz}, \citenamefont
  {Kapitulnik},\ and\ \citenamefont {Eigler}}]{Yazdani1999}%
  \BibitemOpen
  \bibfield  {author} {\bibinfo {author} {\bibfnamefont {A.}~\bibnamefont
  {Yazdani}}, \bibinfo {author} {\bibfnamefont {C.~M.}\ \bibnamefont {Howald}},
  \bibinfo {author} {\bibfnamefont {C.~P.}\ \bibnamefont {Lutz}}, \bibinfo
  {author} {\bibfnamefont {A.}~\bibnamefont {Kapitulnik}}, \ and\ \bibinfo
  {author} {\bibfnamefont {D.~M.}\ \bibnamefont {Eigler}},\ }\bibfield  {title}
  {\enquote {\bibinfo {title} {Impurity-induced bound excitations on the
  surface of
  {${\mathrm{Bi}}_{2}{\mathrm{Sr}}_{2}{\mathrm{CaCu}}_{2}{O}_{8}$}},}\ }\href
  {\doibase 10.1103/PhysRevLett.83.176} {\bibfield  {journal} {\bibinfo
  {journal} {Phys. Rev. Lett.}\ }\textbf {\bibinfo {volume} {83}},\ \bibinfo
  {pages} {176--179} (\bibinfo {year} {1999})}\BibitemShut {NoStop}%
\bibitem [{\citenamefont {Yazdani}(2015)}]{Yazdani2015}%
  \BibitemOpen
  \bibfield  {author} {\bibinfo {author} {\bibfnamefont {A.}~\bibnamefont
  {Yazdani}},\ }\bibfield  {title} {\enquote {\bibinfo {title} {Visualizing
  majorana fermions in a chain of magnetic atoms on a superconductor},}\ }\href
  {\doibase 10.1088/0031-8949/2015/T164/014012} {\bibfield  {journal} {\bibinfo
   {journal} {Physica Scripta}\ }\textbf {\bibinfo {volume} {2015}},\ \bibinfo
  {pages} {014012} (\bibinfo {year} {2015})}\BibitemShut {NoStop}%
\bibitem [{\citenamefont {Schneider}\ \emph {et~al.}(2021)\citenamefont
  {Schneider}, \citenamefont {Beck}, \citenamefont {Posske}, \citenamefont
  {Crawford}, \citenamefont {Mascot}, \citenamefont {Rachel}, \citenamefont
  {Wiesendanger},\ and\ \citenamefont {Wiebe}}]{Wiesendanger2021}%
  \BibitemOpen
  \bibfield  {author} {\bibinfo {author} {\bibfnamefont {L.}~\bibnamefont
  {Schneider}}, \bibinfo {author} {\bibfnamefont {P.}~\bibnamefont {Beck}},
  \bibinfo {author} {\bibfnamefont {T.}~\bibnamefont {Posske}}, \bibinfo
  {author} {\bibfnamefont {D.}~\bibnamefont {Crawford}}, \bibinfo {author}
  {\bibfnamefont {E.}~\bibnamefont {Mascot}}, \bibinfo {author} {\bibfnamefont
  {S.}~\bibnamefont {Rachel}}, \bibinfo {author} {\bibfnamefont
  {R.}~\bibnamefont {Wiesendanger}}, \ and\ \bibinfo {author} {\bibfnamefont
  {J.}~\bibnamefont {Wiebe}},\ }\bibfield  {title} {\enquote {\bibinfo {title}
  {Topological shiba bands in artificial spin chains on superconductors},}\
  }\href {\doibase 10.1038/s41567-021-01234-y} {\bibfield  {journal} {\bibinfo
  {journal} {Nature Physics}\ }\textbf {\bibinfo {volume} {17}},\ \bibinfo
  {pages} {943--948} (\bibinfo {year} {2021})}\BibitemShut {NoStop}%
\bibitem [{\citenamefont {Beck}\ \emph {et~al.}(2021)\citenamefont {Beck},
  \citenamefont {Schneider}, \citenamefont {R{\'o}zsa}, \citenamefont
  {Palot{\'a}s}, \citenamefont {L{\'a}szl{\'o}ffy}, \citenamefont {Szunyogh},
  \citenamefont {Wiebe},\ and\ \citenamefont {Wiesendanger}}]{Beck2021}%
  \BibitemOpen
  \bibfield  {author} {\bibinfo {author} {\bibfnamefont {P.}~\bibnamefont
  {Beck}}, \bibinfo {author} {\bibfnamefont {L.}~\bibnamefont {Schneider}},
  \bibinfo {author} {\bibfnamefont {L.}~\bibnamefont {R{\'o}zsa}}, \bibinfo
  {author} {\bibfnamefont {K.}~\bibnamefont {Palot{\'a}s}}, \bibinfo {author}
  {\bibfnamefont {A.}~\bibnamefont {L{\'a}szl{\'o}ffy}}, \bibinfo {author}
  {\bibfnamefont {L.}~\bibnamefont {Szunyogh}}, \bibinfo {author}
  {\bibfnamefont {J.}~\bibnamefont {Wiebe}}, \ and\ \bibinfo {author}
  {\bibfnamefont {R.}~\bibnamefont {Wiesendanger}},\ }\bibfield  {title}
  {\enquote {\bibinfo {title} {Spin-orbit coupling induced splitting of
  {Yu-Shiba-Rusinov} states in antiferromagnetic dimers},}\ }\href {\doibase
  10.1038/s41467-021-22261-6} {\bibfield  {journal} {\bibinfo  {journal}
  {Nature Communications}\ }\textbf {\bibinfo {volume} {12}},\ \bibinfo {pages}
  {2040} (\bibinfo {year} {2021})}\BibitemShut {NoStop}%
\bibitem [{\citenamefont {Wang}\ \emph {et~al.}(2021)\citenamefont {Wang},
  \citenamefont {Wiebe}, \citenamefont {Zhong}, \citenamefont {Gu},\ and\
  \citenamefont {Wiesendanger}}]{Wang2021}%
  \BibitemOpen
  \bibfield  {author} {\bibinfo {author} {\bibfnamefont {D.}~\bibnamefont
  {Wang}}, \bibinfo {author} {\bibfnamefont {J.}~\bibnamefont {Wiebe}},
  \bibinfo {author} {\bibfnamefont {R.}~\bibnamefont {Zhong}}, \bibinfo
  {author} {\bibfnamefont {G.}~\bibnamefont {Gu}}, \ and\ \bibinfo {author}
  {\bibfnamefont {R.}~\bibnamefont {Wiesendanger}},\ }\bibfield  {title}
  {\enquote {\bibinfo {title} {Spin-polarized yu-shiba-rusinov states in an
  iron-based superconductor},}\ }\href {\doibase
  10.1103/PhysRevLett.126.076802} {\bibfield  {journal} {\bibinfo  {journal}
  {Phys. Rev. Lett.}\ }\textbf {\bibinfo {volume} {126}},\ \bibinfo {pages}
  {076802} (\bibinfo {year} {2021})}\BibitemShut {NoStop}%
\bibitem [{\citenamefont {Schneider}\ \emph {et~al.}(2022)\citenamefont
  {Schneider}, \citenamefont {Beck}, \citenamefont {Neuhaus-Steinmetz},
  \citenamefont {R{\'o}zsa}, \citenamefont {Posske}, \citenamefont {Wiebe},\
  and\ \citenamefont {Wiesendanger}}]{Schneider2022}%
  \BibitemOpen
  \bibfield  {author} {\bibinfo {author} {\bibfnamefont {L.}~\bibnamefont
  {Schneider}}, \bibinfo {author} {\bibfnamefont {P.}~\bibnamefont {Beck}},
  \bibinfo {author} {\bibfnamefont {J.}~\bibnamefont {Neuhaus-Steinmetz}},
  \bibinfo {author} {\bibfnamefont {L.}~\bibnamefont {R{\'o}zsa}}, \bibinfo
  {author} {\bibfnamefont {T.}~\bibnamefont {Posske}}, \bibinfo {author}
  {\bibfnamefont {J.}~\bibnamefont {Wiebe}}, \ and\ \bibinfo {author}
  {\bibfnamefont {R.}~\bibnamefont {Wiesendanger}},\ }\bibfield  {title}
  {\enquote {\bibinfo {title} {Precursors of majorana modes and their
  length-dependent energy oscillations probed at both ends of atomic shiba
  chains},}\ }\href {\doibase 10.1038/s41565-022-01078-4} {\bibfield  {journal}
  {\bibinfo  {journal} {Nature Nanotechnology}\ }\textbf {\bibinfo {volume}
  {17}},\ \bibinfo {pages} {384--389} (\bibinfo {year} {2022})}\BibitemShut
  {NoStop}%
\bibitem [{\citenamefont {Küster}\ \emph {et~al.}(2022)\citenamefont
  {Küster}, \citenamefont {Brinker}, \citenamefont {Hess}, \citenamefont
  {Loss}, \citenamefont {Parkin}, \citenamefont {Klinovaja}, \citenamefont
  {Lounis},\ and\ \citenamefont {Sessi}}]{Richard2022}%
  \BibitemOpen
  \bibfield  {author} {\bibinfo {author} {\bibfnamefont {F.}~\bibnamefont
  {Küster}}, \bibinfo {author} {\bibfnamefont {S.}~\bibnamefont {Brinker}},
  \bibinfo {author} {\bibfnamefont {R.}~\bibnamefont {Hess}}, \bibinfo {author}
  {\bibfnamefont {D.}~\bibnamefont {Loss}}, \bibinfo {author} {\bibfnamefont
  {S.~S.~P.}\ \bibnamefont {Parkin}}, \bibinfo {author} {\bibfnamefont
  {J.}~\bibnamefont {Klinovaja}}, \bibinfo {author} {\bibfnamefont
  {S.}~\bibnamefont {Lounis}}, \ and\ \bibinfo {author} {\bibfnamefont
  {P.}~\bibnamefont {Sessi}},\ }\bibfield  {title} {\enquote {\bibinfo {title}
  {Non-majorana modes in diluted spin chains proximitized to a
  superconductor},}\ }\href {\doibase 10.1073/pnas.2210589119} {\bibfield
  {journal} {\bibinfo  {journal} {Proceedings of the National Academy of
  Sciences}\ }\textbf {\bibinfo {volume} {119}},\ \bibinfo {pages}
  {e2210589119} (\bibinfo {year} {2022})}\BibitemShut {NoStop}%
\bibitem [{\citenamefont {Lo~Conte}\ \emph {et~al.}(2022)\citenamefont
  {Lo~Conte}, \citenamefont {Bazarnik}, \citenamefont {Palot\'as},
  \citenamefont {R\'ozsa}, \citenamefont {Szunyogh}, \citenamefont {Kubetzka},
  \citenamefont {von Bergmann},\ and\ \citenamefont
  {Wiesendanger}}]{Wiesendanger2022}%
  \BibitemOpen
  \bibfield  {author} {\bibinfo {author} {\bibfnamefont {R.}~\bibnamefont
  {Lo~Conte}}, \bibinfo {author} {\bibfnamefont {M.}~\bibnamefont {Bazarnik}},
  \bibinfo {author} {\bibfnamefont {K.}~\bibnamefont {Palot\'as}}, \bibinfo
  {author} {\bibfnamefont {L.}~\bibnamefont {R\'ozsa}}, \bibinfo {author}
  {\bibfnamefont {L.}~\bibnamefont {Szunyogh}}, \bibinfo {author}
  {\bibfnamefont {A.}~\bibnamefont {Kubetzka}}, \bibinfo {author}
  {\bibfnamefont {K.}~\bibnamefont {von Bergmann}}, \ and\ \bibinfo {author}
  {\bibfnamefont {R.}~\bibnamefont {Wiesendanger}},\ }\bibfield  {title}
  {\enquote {\bibinfo {title} {Coexistence of antiferromagnetism and
  superconductivity in {Mn/Nb(110)}},}\ }\href {\doibase
  10.1103/PhysRevB.105.L100406} {\bibfield  {journal} {\bibinfo  {journal}
  {Phys. Rev. B}\ }\textbf {\bibinfo {volume} {105}},\ \bibinfo {pages}
  {L100406} (\bibinfo {year} {2022})}\BibitemShut {NoStop}%
\bibitem [{\citenamefont {Yazdani}\ \emph {et~al.}(2023)\citenamefont
  {Yazdani}, \citenamefont {von Oppen}, \citenamefont {Halperin},\ and\
  \citenamefont {Yacoby}}]{Yacoby2023}%
  \BibitemOpen
  \bibfield  {author} {\bibinfo {author} {\bibfnamefont {A.}~\bibnamefont
  {Yazdani}}, \bibinfo {author} {\bibfnamefont {F.}~\bibnamefont {von Oppen}},
  \bibinfo {author} {\bibfnamefont {B.~I.}\ \bibnamefont {Halperin}}, \ and\
  \bibinfo {author} {\bibfnamefont {A.}~\bibnamefont {Yacoby}},\ }\bibfield
  {title} {\enquote {\bibinfo {title} {Hunting for majoranas},}\ }\href
  {\doibase 10.1126/science.ade0850} {\bibfield  {journal} {\bibinfo  {journal}
  {Science}\ }\textbf {\bibinfo {volume} {380}},\ \bibinfo {pages} {eade0850}
  (\bibinfo {year} {2023})}\BibitemShut {NoStop}%
\bibitem [{\citenamefont {Soldini}\ \emph {et~al.}(2023)\citenamefont
  {Soldini}, \citenamefont {Küster}, \citenamefont {Wagner}, \citenamefont
  {Das}, \citenamefont {Aldarawsheh}, \citenamefont {Thomale}, \citenamefont
  {Lounis}, \citenamefont {Parkin}, \citenamefont {Sessi},\ and\ \citenamefont
  {Neupert}}]{Soldini2023}%
  \BibitemOpen
  \bibfield  {author} {\bibinfo {author} {\bibfnamefont {M.~O.}\ \bibnamefont
  {Soldini}}, \bibinfo {author} {\bibfnamefont {F.}~\bibnamefont {Küster}},
  \bibinfo {author} {\bibfnamefont {G.}~\bibnamefont {Wagner}}, \bibinfo
  {author} {\bibfnamefont {S.}~\bibnamefont {Das}}, \bibinfo {author}
  {\bibfnamefont {A.}~\bibnamefont {Aldarawsheh}}, \bibinfo {author}
  {\bibfnamefont {R.}~\bibnamefont {Thomale}}, \bibinfo {author} {\bibfnamefont
  {S.}~\bibnamefont {Lounis}}, \bibinfo {author} {\bibfnamefont {S.~S.~P.}\
  \bibnamefont {Parkin}}, \bibinfo {author} {\bibfnamefont {P.}~\bibnamefont
  {Sessi}}, \ and\ \bibinfo {author} {\bibfnamefont {T.}~\bibnamefont
  {Neupert}},\ }\bibfield  {title} {\enquote {\bibinfo {title} {Two-dimensional
  shiba lattices as a possible platform for crystalline topological
  superconductivity},}\ }\href {\doibase 10.1038/s41567-023-02104-5} {\bibfield
   {journal} {\bibinfo  {journal} {Nature Physics}\ }\textbf {\bibinfo {volume}
  {19}},\ \bibinfo {pages} {1848--1854} (\bibinfo {year} {2023})}\BibitemShut
  {NoStop}%
\bibitem [{\citenamefont {Subhadarshini}\ \emph {et~al.}(2024)\citenamefont
  {Subhadarshini}, \citenamefont {Pal}, \citenamefont {Chatterjee},\ and\
  \citenamefont {Saha}}]{Subhadarshini2024}%
  \BibitemOpen
  \bibfield  {author} {\bibinfo {author} {\bibfnamefont {M.}~\bibnamefont
  {Subhadarshini}}, \bibinfo {author} {\bibfnamefont {A.}~\bibnamefont {Pal}},
  \bibinfo {author} {\bibfnamefont {P.}~\bibnamefont {Chatterjee}}, \ and\
  \bibinfo {author} {\bibfnamefont {A.}~\bibnamefont {Saha}},\ }\bibfield
  {title} {\enquote {\bibinfo {title} {Multiple topological phase transitions
  unveiling gapless topological superconductivity in magnet/unconventional
  superconductor hybrid platform},}\ }\href {\doibase 10.1063/5.0199275}
  {\bibfield  {journal} {\bibinfo  {journal} {Applied Physics Letters}\
  }\textbf {\bibinfo {volume} {124}},\ \bibinfo {pages} {183102} (\bibinfo
  {year} {2024})}\BibitemShut {NoStop}%
\bibitem [{\citenamefont {Neupert}\ \emph {et~al.}(2016)\citenamefont
  {Neupert}, \citenamefont {Yazdani},\ and\ \citenamefont
  {Bernevig}}]{Neupert2016}%
  \BibitemOpen
  \bibfield  {author} {\bibinfo {author} {\bibfnamefont {T.}~\bibnamefont
  {Neupert}}, \bibinfo {author} {\bibfnamefont {A.}~\bibnamefont {Yazdani}}, \
  and\ \bibinfo {author} {\bibfnamefont {B.~A.}\ \bibnamefont {Bernevig}},\
  }\bibfield  {title} {\enquote {\bibinfo {title} {Shiba chains of scalar
  impurities on unconventional superconductors},}\ }\href {\doibase
  10.1103/PhysRevB.93.094508} {\bibfield  {journal} {\bibinfo  {journal} {Phys.
  Rev. B}\ }\textbf {\bibinfo {volume} {93}},\ \bibinfo {pages} {094508}
  (\bibinfo {year} {2016})}\BibitemShut {NoStop}%
\bibitem [{\citenamefont {Zhang}\ \emph {et~al.}(2019)\citenamefont {Zhang},
  \citenamefont {Cole}, \citenamefont {Wu},\ and\ \citenamefont
  {Das~Sarma}}]{DasSharma_2019}%
  \BibitemOpen
  \bibfield  {author} {\bibinfo {author} {\bibfnamefont {R.-X.}\ \bibnamefont
  {Zhang}}, \bibinfo {author} {\bibfnamefont {W.~S.}\ \bibnamefont {Cole}},
  \bibinfo {author} {\bibfnamefont {X.}~\bibnamefont {Wu}}, \ and\ \bibinfo
  {author} {\bibfnamefont {S.}~\bibnamefont {Das~Sarma}},\ }\bibfield  {title}
  {\enquote {\bibinfo {title} {Higher-order topology and nodal topological
  superconductivity in fe(se,te) heterostructures},}\ }\href {\doibase
  10.1103/PhysRevLett.123.167001} {\bibfield  {journal} {\bibinfo  {journal}
  {Phys. Rev. Lett.}\ }\textbf {\bibinfo {volume} {123}},\ \bibinfo {pages}
  {167001} (\bibinfo {year} {2019})}\BibitemShut {NoStop}%
\bibitem [{\citenamefont {Volpez}\ \emph {et~al.}(2019)\citenamefont {Volpez},
  \citenamefont {Loss},\ and\ \citenamefont {Klinovaja}}]{Dloss_2019}%
  \BibitemOpen
  \bibfield  {author} {\bibinfo {author} {\bibfnamefont {Y.}~\bibnamefont
  {Volpez}}, \bibinfo {author} {\bibfnamefont {D.}~\bibnamefont {Loss}}, \ and\
  \bibinfo {author} {\bibfnamefont {J.}~\bibnamefont {Klinovaja}},\ }\bibfield
  {title} {\enquote {\bibinfo {title} {Second-order topological
  superconductivity in $\ensuremath{\pi}$-junction rashba layers},}\ }\href
  {\doibase 10.1103/PhysRevLett.122.126402} {\bibfield  {journal} {\bibinfo
  {journal} {Phys. Rev. Lett.}\ }\textbf {\bibinfo {volume} {122}},\ \bibinfo
  {pages} {126402} (\bibinfo {year} {2019})}\BibitemShut {NoStop}%
\bibitem [{\citenamefont {Ghosh}\ \emph
  {et~al.}(2021{\natexlab{a}})\citenamefont {Ghosh}, \citenamefont {Nag},\ and\
  \citenamefont {Saha}}]{Ghosh_2021}%
  \BibitemOpen
  \bibfield  {author} {\bibinfo {author} {\bibfnamefont {A.~K.}\ \bibnamefont
  {Ghosh}}, \bibinfo {author} {\bibfnamefont {T.}~\bibnamefont {Nag}}, \ and\
  \bibinfo {author} {\bibfnamefont {A.}~\bibnamefont {Saha}},\ }\bibfield
  {title} {\enquote {\bibinfo {title} {Floquet generation of a second-order
  topological superconductor},}\ }\href {\doibase 10.1103/PhysRevB.103.045424}
  {\bibfield  {journal} {\bibinfo  {journal} {Phys. Rev. B}\ }\textbf {\bibinfo
  {volume} {103}},\ \bibinfo {pages} {045424} (\bibinfo {year}
  {2021}{\natexlab{a}})}\BibitemShut {NoStop}%
\bibitem [{\citenamefont {Li}\ and\ \citenamefont
  {Liu}(2023)}]{Cheng-Cheng_2023}%
  \BibitemOpen
  \bibfield  {author} {\bibinfo {author} {\bibfnamefont {Y.-X.}\ \bibnamefont
  {Li}}\ and\ \bibinfo {author} {\bibfnamefont {C.-C.}\ \bibnamefont {Liu}},\
  }\bibfield  {title} {\enquote {\bibinfo {title} {Majorana corner modes and
  tunable patterns in an altermagnet heterostructure},}\ }\href {\doibase
  10.1103/PhysRevB.108.205410} {\bibfield  {journal} {\bibinfo  {journal}
  {Phys. Rev. B}\ }\textbf {\bibinfo {volume} {108}},\ \bibinfo {pages}
  {205410} (\bibinfo {year} {2023})}\BibitemShut {NoStop}%
\bibitem [{\citenamefont {Liu}\ \emph {et~al.}(2018)\citenamefont {Liu},
  \citenamefont {He},\ and\ \citenamefont {Nori}}]{Franco_2018}%
  \BibitemOpen
  \bibfield  {author} {\bibinfo {author} {\bibfnamefont {T.}~\bibnamefont
  {Liu}}, \bibinfo {author} {\bibfnamefont {J.~J.}\ \bibnamefont {He}}, \ and\
  \bibinfo {author} {\bibfnamefont {F.}~\bibnamefont {Nori}},\ }\bibfield
  {title} {\enquote {\bibinfo {title} {Majorana corner states in a
  two-dimensional magnetic topological insulator on a high-temperature
  superconductor},}\ }\href {\doibase 10.1103/PhysRevB.98.245413} {\bibfield
  {journal} {\bibinfo  {journal} {Phys. Rev. B}\ }\textbf {\bibinfo {volume}
  {98}},\ \bibinfo {pages} {245413} (\bibinfo {year} {2018})}\BibitemShut
  {NoStop}%
\bibitem [{\citenamefont {Wang}\ \emph {et~al.}(2018)\citenamefont {Wang},
  \citenamefont {Liu}, \citenamefont {Lu},\ and\ \citenamefont
  {Zhang}}]{Cheng-Cheng_2018}%
  \BibitemOpen
  \bibfield  {author} {\bibinfo {author} {\bibfnamefont {Q.}~\bibnamefont
  {Wang}}, \bibinfo {author} {\bibfnamefont {C.-C.}\ \bibnamefont {Liu}},
  \bibinfo {author} {\bibfnamefont {Y.-M.}\ \bibnamefont {Lu}}, \ and\ \bibinfo
  {author} {\bibfnamefont {F.}~\bibnamefont {Zhang}},\ }\bibfield  {title}
  {\enquote {\bibinfo {title} {High-temperature majorana corner states},}\
  }\href {\doibase 10.1103/PhysRevLett.121.186801} {\bibfield  {journal}
  {\bibinfo  {journal} {Phys. Rev. Lett.}\ }\textbf {\bibinfo {volume} {121}},\
  \bibinfo {pages} {186801} (\bibinfo {year} {2018})}\BibitemShut {NoStop}%
\bibitem [{\citenamefont {Shiba}(1968)}]{Shiba1968}%
  \BibitemOpen
  \bibfield  {author} {\bibinfo {author} {\bibfnamefont {H.}~\bibnamefont
  {Shiba}},\ }\bibfield  {title} {\enquote {\bibinfo {title} {{Classical Spins
  in Superconductors}},}\ }\href {\doibase 10.1143/PTP.40.435} {\bibfield
  {journal} {\bibinfo  {journal} {Progress of Theoretical Physics}\ }\textbf
  {\bibinfo {volume} {40}},\ \bibinfo {pages} {435--451} (\bibinfo {year}
  {1968})}\BibitemShut {NoStop}%
\bibitem [{\citenamefont {Bernevig}\ \emph {et~al.}(2006)\citenamefont
  {Bernevig}, \citenamefont {Hughes},\ and\ \citenamefont
  {Zhang}}]{science-Bernevig}%
  \BibitemOpen
  \bibfield  {author} {\bibinfo {author} {\bibfnamefont {B.~A.}\ \bibnamefont
  {Bernevig}}, \bibinfo {author} {\bibfnamefont {T.~L.}\ \bibnamefont
  {Hughes}}, \ and\ \bibinfo {author} {\bibfnamefont {S.-C.}\ \bibnamefont
  {Zhang}},\ }\bibfield  {title} {\enquote {\bibinfo {title} {Quantum spin hall
  effect and topological phase transition in hgte quantum wells},}\ }\href
  {\doibase 10.1126/science.1133734} {\bibfield  {journal} {\bibinfo  {journal}
  {Science}\ }\textbf {\bibinfo {volume} {314}},\ \bibinfo {pages} {1757--1761}
  (\bibinfo {year} {2006})}\BibitemShut {NoStop}%
\bibitem [{\citenamefont {König}\ \emph {et~al.}(2007)\citenamefont {König},
  \citenamefont {Wiedmann}, \citenamefont {Brüne}, \citenamefont {Roth},
  \citenamefont {Buhmann}, \citenamefont {Molenkamp}, \citenamefont {Qi},\ and\
  \citenamefont {Zhang}}]{science-zhang}%
  \BibitemOpen
  \bibfield  {author} {\bibinfo {author} {\bibfnamefont {M.}~\bibnamefont
  {König}}, \bibinfo {author} {\bibfnamefont {S.}~\bibnamefont {Wiedmann}},
  \bibinfo {author} {\bibfnamefont {C.}~\bibnamefont {Brüne}}, \bibinfo
  {author} {\bibfnamefont {A.}~\bibnamefont {Roth}}, \bibinfo {author}
  {\bibfnamefont {H.}~\bibnamefont {Buhmann}}, \bibinfo {author} {\bibfnamefont
  {L.~W.}\ \bibnamefont {Molenkamp}}, \bibinfo {author} {\bibfnamefont {X.-L.}\
  \bibnamefont {Qi}}, \ and\ \bibinfo {author} {\bibfnamefont {S.-C.}\
  \bibnamefont {Zhang}},\ }\bibfield  {title} {\enquote {\bibinfo {title}
  {Quantum spin hall insulator state in hgte quantum wells},}\ }\href {\doibase
  10.1126/science.1148047} {\bibfield  {journal} {\bibinfo  {journal}
  {Science}\ }\textbf {\bibinfo {volume} {318}},\ \bibinfo {pages} {766--770}
  (\bibinfo {year} {2007})}\BibitemShut {NoStop}%
\bibitem [{\citenamefont {Hess}\ \emph {et~al.}(2022)\citenamefont {Hess},
  \citenamefont {Legg}, \citenamefont {Loss},\ and\ \citenamefont
  {Klinovaja}}]{DLoss_2022}%
  \BibitemOpen
  \bibfield  {author} {\bibinfo {author} {\bibfnamefont {R.}~\bibnamefont
  {Hess}}, \bibinfo {author} {\bibfnamefont {H.~F.}\ \bibnamefont {Legg}},
  \bibinfo {author} {\bibfnamefont {D.}~\bibnamefont {Loss}}, \ and\ \bibinfo
  {author} {\bibfnamefont {J.}~\bibnamefont {Klinovaja}},\ }\bibfield  {title}
  {\enquote {\bibinfo {title} {Prevalence of trivial zero-energy subgap states
  in nonuniform helical spin chains on the surface of superconductors},}\
  }\href {\doibase 10.1103/PhysRevB.106.104503} {\bibfield  {journal} {\bibinfo
   {journal} {Phys. Rev. B}\ }\textbf {\bibinfo {volume} {106}},\ \bibinfo
  {pages} {104503} (\bibinfo {year} {2022})}\BibitemShut {NoStop}%
\bibitem [{\citenamefont {Benalcazar}\ and\ \citenamefont
  {Cerjan}(2022)}]{Benalcazar_2022}%
  \BibitemOpen
  \bibfield  {author} {\bibinfo {author} {\bibfnamefont {W.~A.}\ \bibnamefont
  {Benalcazar}}\ and\ \bibinfo {author} {\bibfnamefont {A.}~\bibnamefont
  {Cerjan}},\ }\bibfield  {title} {\enquote {\bibinfo {title} {Chiral-symmetric
  higher-order topological phases of matter},}\ }\href {\doibase
  10.1103/PhysRevLett.128.127601} {\bibfield  {journal} {\bibinfo  {journal}
  {Phys. Rev. Lett.}\ }\textbf {\bibinfo {volume} {128}},\ \bibinfo {pages}
  {127601} (\bibinfo {year} {2022})}\BibitemShut {NoStop}%
\bibitem [{\citenamefont {Pal}\ and\ \citenamefont {Ghosh}()}]{pal2024}%
  \BibitemOpen
  \bibfield  {author} {\bibinfo {author} {\bibfnamefont {A.}~\bibnamefont
  {Pal}}\ and\ \bibinfo {author} {\bibfnamefont {A.~K.}\ \bibnamefont
  {Ghosh}},\ }\bibfield  {title} {\enquote {\bibinfo {title} {Multi
  higher-order dirac and weyl semimetals},}\ }\href
  {https://arxiv.org/abs/2408.17152} {\ }\Eprint
  {http://arxiv.org/abs/2408.17152} {arXiv:2408.17152 [cond-mat.mes-hall]}
  \BibitemShut {NoStop}%
\bibitem [{\citenamefont {Luo}\ \emph {et~al.}(2025)\citenamefont {Luo},
  \citenamefont {Li}, \citenamefont {Xiao},\ and\ \citenamefont
  {Wu}}]{Bott_Luo_2025}%
  \BibitemOpen
  \bibfield  {author} {\bibinfo {author} {\bibfnamefont {X.-J.}\ \bibnamefont
  {Luo}}, \bibinfo {author} {\bibfnamefont {J.-Z.}\ \bibnamefont {Li}},
  \bibinfo {author} {\bibfnamefont {M.}~\bibnamefont {Xiao}}, \ and\ \bibinfo
  {author} {\bibfnamefont {F.}~\bibnamefont {Wu}},\ }\bibfield  {title}
  {\enquote {\bibinfo {title} {Characterization of higher-order topological
  superconductors using bott indices},}\ }\href {\doibase
  10.1103/PhysRevB.111.184516} {\bibfield  {journal} {\bibinfo  {journal}
  {Phys. Rev. B}\ }\textbf {\bibinfo {volume} {111}},\ \bibinfo {pages}
  {184516} (\bibinfo {year} {2025})}\BibitemShut {NoStop}%
\bibitem [{\citenamefont {Ghosh}\ \emph
  {et~al.}(2021{\natexlab{b}})\citenamefont {Ghosh}, \citenamefont {Nag},\ and\
  \citenamefont {Saha}}]{Ghosh_2021b}%
  \BibitemOpen
  \bibfield  {author} {\bibinfo {author} {\bibfnamefont {A.~K.}\ \bibnamefont
  {Ghosh}}, \bibinfo {author} {\bibfnamefont {T.}~\bibnamefont {Nag}}, \ and\
  \bibinfo {author} {\bibfnamefont {A.}~\bibnamefont {Saha}},\ }\bibfield
  {title} {\enquote {\bibinfo {title} {Floquet second order topological
  superconductor based on unconventional pairing},}\ }\href {\doibase
  10.1103/PhysRevB.103.085413} {\bibfield  {journal} {\bibinfo  {journal}
  {Phys. Rev. B}\ }\textbf {\bibinfo {volume} {103}},\ \bibinfo {pages}
  {085413} (\bibinfo {year} {2021}{\natexlab{b}})}\BibitemShut {NoStop}%
\bibitem [{\citenamefont {Fu}\ and\ \citenamefont {Kane}(2007)}]{FuKane_2007}%
  \BibitemOpen
  \bibfield  {author} {\bibinfo {author} {\bibfnamefont {L.}~\bibnamefont
  {Fu}}\ and\ \bibinfo {author} {\bibfnamefont {C.~L.}\ \bibnamefont {Kane}},\
  }\bibfield  {title} {\enquote {\bibinfo {title} {Topological insulators with
  inversion symmetry},}\ }\href {\doibase 10.1103/PhysRevB.76.045302}
  {\bibfield  {journal} {\bibinfo  {journal} {Phys. Rev. B}\ }\textbf {\bibinfo
  {volume} {76}},\ \bibinfo {pages} {045302} (\bibinfo {year}
  {2007})}\BibitemShut {NoStop}%
\bibitem [{\citenamefont {Jackiw}\ and\ \citenamefont
  {Rebbi}(1976)}]{Jackiw_Rebbi_1976}%
  \BibitemOpen
  \bibfield  {author} {\bibinfo {author} {\bibfnamefont {R.}~\bibnamefont
  {Jackiw}}\ and\ \bibinfo {author} {\bibfnamefont {C.}~\bibnamefont {Rebbi}},\
  }\bibfield  {title} {\enquote {\bibinfo {title} {Solitons with fermion number
  \textonehalf{}},}\ }\href {\doibase 10.1103/PhysRevD.13.3398} {\bibfield
  {journal} {\bibinfo  {journal} {Phys. Rev. D}\ }\textbf {\bibinfo {volume}
  {13}},\ \bibinfo {pages} {3398--3409} (\bibinfo {year} {1976})}\BibitemShut
  {NoStop}%
\bibitem [{\citenamefont {Chatzopoulos}\ \emph {et~al.}(2021)\citenamefont
  {Chatzopoulos}, \citenamefont {Cho}, \citenamefont {Bastiaans}, \citenamefont
  {Steffensen}, \citenamefont {Bouwmeester}, \citenamefont {Akbari},
  \citenamefont {Gu}, \citenamefont {Paaske}, \citenamefont {Andersen},\ and\
  \citenamefont {Allan}}]{Chatzopoulos2021}%
  \BibitemOpen
  \bibfield  {author} {\bibinfo {author} {\bibfnamefont {D.}~\bibnamefont
  {Chatzopoulos}}, \bibinfo {author} {\bibfnamefont {D.}~\bibnamefont {Cho}},
  \bibinfo {author} {\bibfnamefont {K.~M.}\ \bibnamefont {Bastiaans}}, \bibinfo
  {author} {\bibfnamefont {G.~O.}\ \bibnamefont {Steffensen}}, \bibinfo
  {author} {\bibfnamefont {D.}~\bibnamefont {Bouwmeester}}, \bibinfo {author}
  {\bibfnamefont {A.}~\bibnamefont {Akbari}}, \bibinfo {author} {\bibfnamefont
  {G.}~\bibnamefont {Gu}}, \bibinfo {author} {\bibfnamefont {J.}~\bibnamefont
  {Paaske}}, \bibinfo {author} {\bibfnamefont {B.~M.}\ \bibnamefont
  {Andersen}}, \ and\ \bibinfo {author} {\bibfnamefont {M.~P.}\ \bibnamefont
  {Allan}},\ }\bibfield  {title} {\enquote {\bibinfo {title} {Spatially
  dispersing yu-shiba-rusinov states in the unconventional superconductor
  fete0.55se0.45},}\ }\href {\doibase 10.1038/s41467-020-20529-x} {\bibfield
  {journal} {\bibinfo  {journal} {Nature Communications}\ }\textbf {\bibinfo
  {volume} {12}},\ \bibinfo {pages} {298} (\bibinfo {year} {2021})}\BibitemShut
  {NoStop}%
\bibitem [{\citenamefont {Medvedev}\ \emph {et~al.}(2009)\citenamefont
  {Medvedev}, \citenamefont {McQueen}, \citenamefont {Troyan}, \citenamefont
  {Palasyuk}, \citenamefont {Eremets}, \citenamefont {Cava}, \citenamefont
  {Naghavi}, \citenamefont {Casper}, \citenamefont {Ksenofontov}, \citenamefont
  {Wortmann},\ and\ \citenamefont {Felser}}]{Medvedev2009}%
  \BibitemOpen
  \bibfield  {author} {\bibinfo {author} {\bibfnamefont {S.}~\bibnamefont
  {Medvedev}}, \bibinfo {author} {\bibfnamefont {T.~M.}\ \bibnamefont
  {McQueen}}, \bibinfo {author} {\bibfnamefont {I.~A.}\ \bibnamefont {Troyan}},
  \bibinfo {author} {\bibfnamefont {T.}~\bibnamefont {Palasyuk}}, \bibinfo
  {author} {\bibfnamefont {M.~I.}\ \bibnamefont {Eremets}}, \bibinfo {author}
  {\bibfnamefont {R.~J.}\ \bibnamefont {Cava}}, \bibinfo {author}
  {\bibfnamefont {S.}~\bibnamefont {Naghavi}}, \bibinfo {author} {\bibfnamefont
  {F.}~\bibnamefont {Casper}}, \bibinfo {author} {\bibfnamefont
  {V.}~\bibnamefont {Ksenofontov}}, \bibinfo {author} {\bibfnamefont
  {G.}~\bibnamefont {Wortmann}}, \ and\ \bibinfo {author} {\bibfnamefont
  {C.}~\bibnamefont {Felser}},\ }\bibfield  {title} {\enquote {\bibinfo {title}
  {Electronic and magnetic phase diagram of $\beta$-fe1.01se with
  superconductivity at 36.7{\thinspace}k under pressure},}\ }\href {\doibase
  10.1038/nmat2491} {\bibfield  {journal} {\bibinfo  {journal} {Nature
  Materials}\ }\textbf {\bibinfo {volume} {8}},\ \bibinfo {pages} {630--633}
  (\bibinfo {year} {2009})}\BibitemShut {NoStop}%
\bibitem [{\citenamefont {Kamihara}\ \emph {et~al.}(2008)\citenamefont
  {Kamihara}, \citenamefont {Watanabe}, \citenamefont {Hirano},\ and\
  \citenamefont {Hosono}}]{Kamihara2008}%
  \BibitemOpen
  \bibfield  {author} {\bibinfo {author} {\bibfnamefont {Y.}~\bibnamefont
  {Kamihara}}, \bibinfo {author} {\bibfnamefont {T.}~\bibnamefont {Watanabe}},
  \bibinfo {author} {\bibfnamefont {M.}~\bibnamefont {Hirano}}, \ and\ \bibinfo
  {author} {\bibfnamefont {H.}~\bibnamefont {Hosono}},\ }\bibfield  {title}
  {\enquote {\bibinfo {title} {Iron-based layered superconductor la[o1-xfx]feas
  (x = 0.05-0.12) with tc = 26 k},}\ }\href {\doibase 10.1021/ja800073m}
  {\bibfield  {journal} {\bibinfo  {journal} {Journal of the American Chemical
  Society}\ }\textbf {\bibinfo {volume} {130}},\ \bibinfo {pages} {3296--3297}
  (\bibinfo {year} {2008})}\BibitemShut {NoStop}%
\bibitem [{\citenamefont {Zhang}\ \emph {et~al.}(2018)\citenamefont {Zhang},
  \citenamefont {Yaji}, \citenamefont {Hashimoto}, \citenamefont {Ota},
  \citenamefont {Kondo}, \citenamefont {Okazaki}, \citenamefont {Wang},
  \citenamefont {Wen}, \citenamefont {Gu}, \citenamefont {Ding},\ and\
  \citenamefont {Shin}}]{Peng2018}%
  \BibitemOpen
  \bibfield  {author} {\bibinfo {author} {\bibfnamefont {P.}~\bibnamefont
  {Zhang}}, \bibinfo {author} {\bibfnamefont {K.}~\bibnamefont {Yaji}},
  \bibinfo {author} {\bibfnamefont {T.}~\bibnamefont {Hashimoto}}, \bibinfo
  {author} {\bibfnamefont {Y.}~\bibnamefont {Ota}}, \bibinfo {author}
  {\bibfnamefont {T.}~\bibnamefont {Kondo}}, \bibinfo {author} {\bibfnamefont
  {K.}~\bibnamefont {Okazaki}}, \bibinfo {author} {\bibfnamefont
  {Z.}~\bibnamefont {Wang}}, \bibinfo {author} {\bibfnamefont {J.}~\bibnamefont
  {Wen}}, \bibinfo {author} {\bibfnamefont {G.~D.}\ \bibnamefont {Gu}},
  \bibinfo {author} {\bibfnamefont {H.}~\bibnamefont {Ding}}, \ and\ \bibinfo
  {author} {\bibfnamefont {S.}~\bibnamefont {Shin}},\ }\bibfield  {title}
  {\enquote {\bibinfo {title} {Observation of topological superconductivity on
  the surface of an iron-based superconductor},}\ }\href {\doibase
  10.1126/science.aan4596} {\bibfield  {journal} {\bibinfo  {journal}
  {Science}\ }\textbf {\bibinfo {volume} {360}},\ \bibinfo {pages} {182--186}
  (\bibinfo {year} {2018})}\BibitemShut {NoStop}%
\end{thebibliography}%
	
\end{document}